\documentclass[fleqn,usenatbib]{mnras}
\usepackage{newtxtext,newtxmath}
\let\Bbbk\relax 
\usepackage[T1]{fontenc}
\DeclareRobustCommand{\VAN}[3]{#2}
\let\VANthebibliography\thebibliography
\def\thebibliography{\DeclareRobustCommand{\VAN}[3]{##3}\VANthebibliography}
\usepackage{graphicx}	
\usepackage{amsmath}	
\usepackage{braket}
\usepackage{caption}
\usepackage{subcaption}
\usepackage{float}
\usepackage{wrapfig}
\usepackage{siunitx}
\usepackage{wasysym}
\usepackage{siunitx}
\usepackage{amssymb}     
\usepackage{hyperref}    
\usepackage{xcolor}      
\usepackage{academicons} 

\title[]{Radial velocity analysis of stars with debris discs}

\newcommand{\orcid}[1]{\href{https://orcid.org/#1}{\includegraphics[width=8pt]{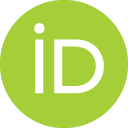}}}

\author[Bisht $\&$ Jones (2024)]{
Deepak Bisht \orcid{0009-0006-7546-5402},$^{1,}$$^{2}$\thanks{E-mail: bishtd78@gmail.com},
Hugh R. A. Jones$^{3}$\thanks{E-mail: h.r.a.jones@herts.ac.uk}
\\
$^{1}$Space Research Institute, Austrian Academy of Sciences, Schmiedlstrasse 6, A-8042 Graz, Austria\\
$^{2}$Indian Institute Of Science Education and Research Berhampur, Odisha-760010, India\\
$^{3}$Centre for Astrophysics Research, University of Hertfordshire, Hatfield, Hertfordshire, AL10 9AB, UK\\
}

\date{}

\pubyear{2024}

\begin{document}
\label{firstpage}
\pagerange{\pageref{firstpage}--\pageref{lastpage}}
\maketitle
\begin{abstract}
This study aims to identify potential exoplanet signals from nearby stars with resolved debris discs. However, the high activity of many stars with debris discs limits the detection of periodic signals. Our study is constrained to a sample of 29 stars that have appropriate radial velocity data and debris disc measurements sufficient to resolve their inclination. Our results confirm and update previous findings for exoplanets around HD~10647, HD~115617, HD~69830, GJ~581, HD~22049, and HD~142091, and we identify long-term activity signals around HD~207129 and HD~202628. We utilize the inclination angles of the debris discs, assuming co-planarity between debris disc and exoplanet orbit, to determine the ``disc-aligned'' masses of radial velocity exoplanets in this study. The ``disc-aligned'' masses of HD 69830 b, HD 69830 c, and 61 Vir b suggests that they may be classified as `hot' or `warm' Jupiters and so might be nearby examples of planets that have undergone recent type-II disc migration.  
\end{abstract}




\section{Introduction}
The discovery of planets orbiting stars outside our Solar System has transformed our understanding of the cosmos. Exploring these exoplanets is vital for unraveling the diversity of planetary systems. This research is driven by the quest to identify and characterize exoplanets within resolved debris discs surrounding nearby stars. Debris discs are circumstellar structures composed of dust and debris that encircle stars. These discs consist of small particles ranging from micrometers to centimeters and are believed to originate from the collision and erosion of planetesimals, remnants of planetary building blocks, during the early stages of planetary system formation \citep{chambers2004planetary}. \\
\\
Debris discs are commonly observed around both young and mature stars, offering valuable insights into the evolutionary processes of planetary systems \citep{krivov2010debris}. These discs probably have two distinct phases.  Initially, a protoplanetary disc consisting of gas and dust develops around a young star, followed by the gradual dissipation of the protoplanetary material after planet formation. This process results in the formation of a debris disc, which is composed of planets and residual planetesimal material \citep{wyatt2016insights}. As a result, debris discs present an opportunity to examine mature planetary systems that could potentially harbor additional exoplanets \citep{pearce2022planet, perez2019upper, milli2017discovery}.\\
\\
The detection of debris discs and the material within them presents challenges due to the variability within the debris disc as well as that of the surrounding environment \citep{beichman2006new}. Various techniques have been developed to detect and study debris discs, including direct imaging, particularly in the infrared and at longer wavelengths \citep{kilic2005excess}. Direct high-resolution imaging of the disc structure enables detailed analysis of its properties \citep{schneider2014probing}. Space-based telescopes, like the Spitzer Space Telescope and the Herschel Space Observatory, along with ground-based observatories, such as ALMA, VLT, and the Keck Observatory, have been instrumental in detecting and analysing debris discs \citep{macgregor2022alma, beuzit2019sphere, booth2016resolving, pilbratt2010herschel,  koerner2010new, werner2004spitzer,  nelson1985design}.\\
\\
After \citet{mayor1995jupiter} detected the first planet around a Sun-like star using the radial velocity (RV) method, over a thousand exoplanets have been confirmed using this technique \citep{akeson2013nasa}. Nevertheless, RV surveys encounter challenges, such as the necessity for high-precision measurements and the influence of stellar activity \citep{zechmeister2018spectrum, jones2006high}. Additionally, accurately determining the inclination angle of a planet's orbit relative to the plane of the sky can be instrumental in RV surveys \citep{zurlo2018imaging, rivera2010lick}. This angle can provide the true mass of a planet, yet this is not routinely available unless the planet also transits the star to provide precise knowledge of the planet's orbital parameters. In recent times, advancements in instrumentation and measuring techniques have enabled the measurement of orbital angles through astrometry. Various studies have utilized astrometric data and the RV method \citep{xiao2023masses, yahalomi2023detecting, barbato2023coralie,  kiefer2021determining, llop2021constraining, li2021precise, xuan2020mutual, benedict2006extrasolar}  to determine the true masses of planets.\\
\\
Planetary systems are recognized to form from discs surrounding young stars. Consequently, it is assumed that constituents of exoplanetary systems should exhibit a unified angular momentum direction. This implies that planets and debris discs should orbit in alignment, following the same direction, and in the same plane as the stellar equator \citep{kennedy2013star}. Investigations have largely indicated co-planarity or near-co-planarity between debris discs and planetary orbits. In the Solar System, for instance, the orbital planes of the planets exhibit a close alignment with each other, as well as with the Kuiper belt. This observation suggests a scenario of planet formation within a flat protoplanetary disc. There are a number of well-studied systems in the literature; for example, the orbit of $\beta$ Pic b is determined to be nearly aligned with the disc mid-plane, with a tilt angle of approximately 2.4$^\circ$\citep{matra2019kuiper}. Additionally, it closely aligns with both the stellar spin axis \citep{kraus2020spin} and the orbit of the inner planet \citep{nowak2020direct}. In HR 8799 \citep{gozdziewski2018orbital} and HD 82943 \citep{kennedy2013star}, the planets exhibit consistency in alignment with both the debris disc and the stellar spin axis. This alignment is inferred from similarities in their sky-projected inclinations. AU Mic was discovered to host a short-period transiting planet that is in alignment with its debris disc and stellar spin axis \citep{palle2020transmission, plavchan2020planet}. Additionally, \citet{pearce2024effect} proposes that the mean motion of a planet in a coplanar disc excites debris eccentricity but does not influence the inclination angle of the debris disc. It is worth noting that existing evidence, at least for planet-hosting binaries, favors low mutual inclinations between planetary and stellar orbits \citep{dupuy2022orbital} and, in contrast to determining a planet's orbital inclination, the inclination angle of a debris disc can often be deduced relatively easily from its shape and orientation relative to the star's equatorial plane \citep{watson2011alignment}. For our analysis, we assume a mutual inclination for the disc and planetary orbits \citep{kennedy2013star} and calculate the Disc Aligned mass ($M_{DA}$).\\
\\
It is crucial to acknowledge that there are systems where the disc and planetary orbits are misaligned \citep{xuan2020mutual}. Theoretical work has indicated that these misalignments could result from misaligned stellar or planetary-mass companions within the system \citep{10.1093/mnras/stz346, zhu2019inclined}. Secular perturbations induced by companions can lead to precession in the orbits of planetesimals within the disc. This precession might cause a deviation in the disc's mid-plane from its initial orientation. For example, \citet{xiang2013interaction} found significant changes in the form of a disc that was produced by the interaction with larger planet masses, such as visible wrapping in the inner parts, and the difference between the inclination of the inner disc and outer disc is up to 10-20 degrees. However, a disc that was initially misaligned with a planet could, over time and multiple precession periods, become aligned with it due to stronger frictional interactions \citep{xiang2013interaction}. This realignment process may reshape the disc as it unfolds \citep{poblete2023self, pearce2014dynamical, kennedy201299, mouillet1997planet}. So the alignment of planetary systems can serve as an indicator of either primordial conditions or the outcome of subsequent dynamical interactions, contingent upon the system's age. Additionally, high-resolution infrared imaging finds that plenty of discs have non-axisymmetric features in scattered light. We can see that in the closest example TW Hydra, these changes happen on a timescale of less than 20 years \citep{debes2023surprising}, but it is unknown how common these are.\\
\\
After introducing our methodology in Section 2, in Section 3 our analysis is structured into two distinct sections. In Section 3.1, we employ the inclination values of debris discs to determine the $M_{DA}$ of RV-detected exoplanets that have been identified around HD 10647, HD 115617, HD 69830, GJ 581, HD 22049, and HD 142091. In Section 3.2, we focus on identifying debris disc stars that have not exhibited any prior RV-detected exoplanets. In Section 4 we conclude with a discussion of our findings.

\section{Methodology}
Our main source to identify stars with debris discs was the catalog of circumstellar discs \footnote{\url{https://www.circumstellardisks.org/}} \citep{2007lyot.confS..47M}, complemented by supplementary sources from the literature \citep{hengst2020multi, macgregor2017complete, tanner2015stellar,  eiroa2013dust, metchev2005adaptive, maldonado2010spectroscopy}. Table \ref{main} is our compilation of stars with resolved debris discs, which have been the subject of RV monitoring programs. Specifically, they were required to possess more than 50 RV data points and exhibit temporal coverage exceeding 1000 days. Where available, disc inclination angles were sourced from the literature. For specific cases, such as HIP 29271 and HIP 85235, where inclination angle data are not in the literature, a direct geometric approach was employed to calculate the inclination angle. It was assumed that the debris discs are circular with zero eccentricity \citep{mustill2009debris} and the difference in the lengths of the major and minor axes arises from the disc inclination, prompting the calculation of inclination angles using the measurements of the major and minor axes from \citet{eiroa2013dust}. It should be noted that this assumption may not always hold true, as evidenced by cases of warped discs and non-zero eccentricities \citep{sakai2019warped, bouvier1999magnetospheric}.\\
\\
The RV data was found from searches across various archives, including the European Southern Observatory (ESO) archive \citep{esoarchive}, the Data Analysis Center for Exoplanets (DACE), Exostriker \citep{trifonov2019exo}, and the HARPS-RVBank archive \citep{trifonov2020public}. This provided the RV data for the specific stars as listed in Table \ref{main}. We extracted the HARPS RV data from the Data Reduction Software (DRS) pipeline within the ESO archive as well as from TERRA pipeline for high activity stars. Importantly, an instrumental update to HARPS in May 2015 led to a discernible jump in the RV data. We utilize the correlation established by \citet{trifonov2020public}, between the RV jump and the spectral type of the observed stars in order to make adjustments to the RV data. After acquiring the RV data, we cross-referenced our findings with sources such as \citet{exoplanetCatalog} and \citet{exoplanetEncyclopedia} to determine whether any exoplanets had been previously identified around these stars using the RV technique.\\
\\
We conducted our analysis on eight stars known to host RV-detected exoplanets: GJ 581, HD 22049, HD 38858, HD 69830, HD 115617, HD 142091, HD 10700, and HD 10647. For our investigation, we utilized a combined RV dataset for these stars. Notably, for some stars $-$ HD 22049, HD 10647, HD 115617, and HD 142091 $-$ we managed to assemble RV datasets larger than those used in previous studies and so provide refined orbital parameters for these stars. Additionally, by assuming the co-planar alignment of both the debris disc and planetary orbits, we calculate the $M_{DA}$ of the RV-detected exoplanets orbiting around them. In the cases of GJ 581 and HD 69830, we were unable to obtain any new RV data beyond what had been previously employed in the literature. For these two stars, we use the available parameters from the literature, and additionally incorporate the inclination angle of the debris disc to deduce $M_{DA}$ for their exoplanets. After this, we conducted a search for exoplanet signals using the RV analysis on the remaining 21 stars, as listed in Table \ref{main}.\\

\begin{table*}
    \centering
    \begin{tabular}{cccccc}
        \hline
        Star&i$^\circ$&Type&RV-detected exoplanets&$RV_{RMS}$ (m/s)&References\\
        \hline
        HD 10647&$76.7\pm1.0$ &F9V&1&11.61&\cite{lovell2021high, marmier2013coralie}\\
        GJ 581 & $50 \pm 20$&M3V&3& 9.38&\cite{reyle202110, trifonov2018carmenes, lestrade2012debris}\\
        HD 142091&$58\pm1$&K0III&1&7.55&\cite{bonsor2013spatially, baines2013navy, sato2012substellar}\\
        HD 22049& $78.26^{+28.60}_{21.56}$ & K2V&1 & 5.89 & \cite{llop2021constraining, greaves2013alignment, wenger2000simbad}\\
        HD 115617 & $77 \pm 4$&G7V&3& 4.39&\cite{laliotis2023doppler, wyatt2012herschel, wenger2000simbad}\\
        HD 69830& $13^{+27}_{-13}$& G8V&3& 3.78&\citet{tanner2015stellar, simpson2010rotation, payne2008dynamical}\\
        HD 38858& 48& G2V &1&	2.59&\cite{flores2018hd, kennedy2015kuiper}\\
        HD 10700&$35\pm10$&G8V&4&1.92&\cite{feng2017color, lawler2014debris, keenan1989perkins}\\
        HD 71155 & 30 & A0V &$-$&	642&\citet{booth2013resolved} \\
        HD 218396 &$31\pm3$&A5V&$-$&610&\cite{pearce2022planet, gray1999hr}\\
        HD 106906&$84.3-85.0$&F5V&$-$&483&\cite{kalas2015direct}\\
        HD 188228 & $50\pm 20$ & A0V &$-$&339&\citet{pearce2022planet, booth2013resolved} \\
        HD 39060&$85.3^{+0.3}_{-0.2}$&A6V&$-$&303&\cite{millar2015beta, gray2006contributions}\\
        HD 95086 & $23.3\pm5.6$ & A8III &$-$&229&\citet{moor2015stirring} \\
        HD 216956 &$65.6\pm0.3$ & A3V &$-$& 227&\citet{macgregor2017complete} \\
        HD 172555&14&A7V &$-$&217& \cite{engler2018detection}\\
        HD 197481 & 89 & M1V &$-$& 188&\cite{torres2006search, metchev2005adaptive}\\
        HD 141943&85&G2V&$-$&170&\cite{soummer2014five}\\
        HD 202917&$68.6\pm1.5$&G7V&$-$&123&\cite{schneider2016deep}\\
        HD 159492 & $40\pm14$  & A5V&$-$ &123&\cite{morales2016herschel}\\
        HD 206893 & $40 \pm 10$  &F5V &$-$&120&\cite{hinkley2023direct, milli2017discovery}\\
        HD 181327& $30.0 \pm 0.5$&F5V&$-$&15.74&\cite{pearce2022planet, schneider2006discovery} \\
        HD 53143&45&G9V&$-$&15.06&\cite{schneider2014probing}\\
        HD 30495 & $51\pm 10$&G2V&$-$&12.25&\cite{maldonado2010spectroscopy, wenger2000simbad}\\
        HD 202628&$57.4\pm0.4$&G2V&$-$&8.16&\cite{faramaz2019scattered}\\
        HD 48682& $67.5\pm 4.2$ &G0V&$-$&5.48&\cite{hengst2020multi}\\
        HD 207129 &$60\pm3$& G2V &$-$&3.87&\cite{krist2010hst}\\
        HIP 29271 & $56.44\pm0.23$&G5V&$-$&3.73&\cite{eiroa2013dust, refregier2008summary}\\
        HIP 85235 & $66.93\pm0.04$&K0V&$-$& 2.75&\cite{eiroa2013dust, refregier2008summary}\\        
        \hline
    \end{tabular}
    \caption{ Compilation of stars with resolved debris discs, inclination angle data, more than 50 RV data points, and temporal coverage exceeding 1000 days. We compile stars that have been identified to host exoplanets in existing literature through the use of the RV method. These stars are then sorted in a descending sequence according to their $RV_{RMS}$ values. Likewise, for all remaining stars, we follow a similar approach, arranging them based on their $RV_{RMS}$ values in descending order as well. Column 1 lists the selected stars, Column 2 displays the inclination angle of their debris discs, and Column 3 corresponds to the spectral type of each star. Column 4 indicates the number of previously detected exoplanets around these stars using RV analysis. Column 5 denotes the $RV_{RMS}$ prior to the fitting of planetary models. Column 6 provides references for the inclinations of the debris discs, the spectral classifications of the stars, and the planetary status.}
    \label{main}
\end{table*}

Our aim is to identify any potential periodic signals indicating the presence of a companion. For this purpose, we plotted Lomb-Scargle (LS) periodograms using DACE. Before constructing the LS periodogram, we factored in the influence of both stellar jitter and instrumental jitter, focusing solely on this aspect in the initial step of our analysis. Later, we utilized Markov Chain Monte Carlo (MCMC) to fit the jitter as a free parameter. To determine the stellar jitter values, we conducted a review of relevant literature to gather any documented values associated with the star under study. In cases where such values were unavailable in the literature, we employed the methodology outlined in \citet{isaacson2010chromospheric} to calculate the stellar jitter. Additionally, we considered the reported instrumental precision for various instruments: 2 m/s for APF \citep{vogt2014apf}, 5 m/s for HAMILTON \citep{tal2019correcting}, 1 m/s for HIRES \citep{tal2019correcting}, 1 m/s for HARPS \citep{jenkins2015observed}, 5 m/s for CORALIE \citep{segransan2010coralie}, and 3 m/s for UCLES \citep{jenkins2015observed}. In our LS periodogram analysis, we considered signals to be significant only if they exhibited a False Alarm Probability (FAP) smaller than 10$\%$ \citep{vanderplas2018understanding, stempels2007periodic}. Our main goal was to differentiate between peaks indicative of exoplanet signals and those arising from stellar activity.\\
\\
After detecting a potential periodic signal, we examined the rotational period of the star. As suggested by \citet{vanderburg2016radial} and \citet{oshagh2013effect}, periodic signals at the rotational period could potentially arise from stellar modulation, indicating stellar activity rather than an exoplanet signal. We ascertained stellar rotational periods from the literature and where possible, refined these values through the utilization of TESS (Transiting Exoplanet Survey Satellite) lightcurve data \citep{2024MNRAS.529.4442S, ricker2010transiting}. Subsequently, after neglecting this signal associated with stellar activity in our periodogram, we proceeded to identify other significant peaks that could potentially indicate exoplanet signals. In certain instances, chromospheric activity was detected at the designated period. Chromospheric activity has the potential to induce both photometric and spectroscopic fluctuations, which within the fitting error resemble exoplanet signals \citep{martinez2010chromospheric, rajpaul2015gaussian}. Therefore for checking the activity signal we perform an analysis utilizing the S-Index and BIS activity indicators. If we detect the signal in the activity indicators, we classify them as possible activity signals rather than exoplanet signals \citep{anna2022impact, toledo2019stellar}. Once a period was selected as a potential exoplanet signal, we conducted an analysis to identify aliases of these periods using DACE. Alias peaks emerge due to the periodicity of observations and can surface even in the absence of authentic periodic signals within the data \citep{VanderPlas2018}. Recognizing these alias peaks as artifacts stemming from the inherent periodic nature of observations \citep{vogt2010lick}, we disregard them due to their limited relevance in confirming the presence of exoplanets.\\
\\
Once a potential exoplanet signal was determined using the periodogram, we utilized DACE to fit a Keplerian model to the RV data, thereby extracting parameters of the planet's orbit \citep{butler2017lces, pepe2013earth}. We then use these parameters as priors and conduct MCMC \citep{diaz2016sophie, diaz2014pastis} simulation to determine the orbital parameters, mass, and orbital distance of the identified planets.  For each MCMC simulation, we performed 1000000 iterations. Additionally, we also compared these results with the results from Radvel \citep{fulton2018radvel} there was no significant change so we used DACE for the RV MCMC analysis. The parameters obtained from the fitting process encompassed the planet's mass M$_{\rm P}$, argument of Periastron ($\omega$), orbital period (P), eccentricity (e), time of Periastron Passage ($T_P$), orbital distance, and velocity amplitude (K). Subsequently, in order to visually validate the model, we employed a phase-folding approach on both the RV data and the model. This procedure proved beneficial in mitigating uncertainties arising from irregularities in data sampling \citep{grunblatt2015determining}. We assessed the goodness of fit by conducting a reduced chi-squared test ($\chi^2_r$) \citep{pepe2013earth} and the Bayesian Information Criterion (BIC) for each model and compared it with the BIC of a straight-line model and calculated the $\Delta$BIC.

\section{Data Analysis}
Among the stars mentioned in table \ref{main}, HD 38858 exhibited two prominent signals within our dataset at periods of 3879.41 and 417.17 days. Further investigation revealed that the 417.17-day signal was an alias of the 3879.14-day signal, likely arising due to a one-year periodicity. When fitting a Keplerian model to the 3879.14-day signal, an orbit with a high eccentricity (e $\approx$ 0.55) was suggested. However, upon closer examination, the 3879.41-day signal was identified as chromospheric activity according to \citet{flores2018hd}, rather than being indicative of an exoplanet. In our analysis, we also examined the S-Index indicator and identified a peak near the 3879 day period, consistent with the findings of \citet{flores2018hd}, who similarly observed the signal in both the S-Index and I$_{\alpha}$ indicators and confirmed it as indicative of chromosphere activity.\\
\\
HD 10700 has undergone extensive monitoring by HARPS. While numerous planet candidates have been identified, there remains a lack of consensus among various analyses \citep{coffinet2019new, feng2017color, tuomi2013signals}. In our investigation, we integrated a total of 10,603 RV data points acquired from the HARPS instrument, along with an additional 803 data points from HIRES. However, this dataset spans a considerable duration of 6,350 days, and the periodic signals detected within it are relatively weak. Moreover, \citet{allart2022automatic} found that the RV signature of micro-tellurics in HD 10700 is up to 58 cm/s, similar in magnitude to the RV amplitudes of the putative exoplanets. They caution that interpreting small-amplitude exoplanet signals requires consideration of micro-tellurics. However, such analysis is beyond the scope of this study. Therefore, for the purpose of this work, HD 10700 is considered a star without RV-detected planetary companions.\\
\\
Subsequently, we observed that the stars HD 39060, HD 172555, HD 106906, HD 141943, HD 202917, HD 71155, HD 95086, HD 159492, HD 197481, HD 206893, HD 216956, HD 218396, and HD 188228 stand out due to their significantly elevated $RV_{RMS}$ values of greater than 100. We note that for these stars, the DRS pipeline produces significantly larger RMS residuals and so we also used TERRA for data reduction, following the methodology outlined by \citet{anglada2012harps}. Upon comparing the resulting $RV_{RMS}$ value with the literature, we observed that the $RV_{RMS}$ of HD 188228 was 339 m/s, similar to the value of 319 m/s found by the work of \citet{grandjean2020harps} specifically focused on hot stars. Therefore, we proceeded with the $RV_{RMS}$ based on TERRA reduction. Upon conducting an RV analysis of these stars, we did not identify any significant periodic signal in 13 out of 14 stars. It is evident that some of the stars with high values of RV$_{RMS}$ already harbor directly imaged exoplanets. However, planets around five of the stars exhibit long orbital periods. Due to the absence of RV data covering such extended temporal ranges, we are unable to detect these planets in our data. However, for HD 206893, \citet{hinkley2023direct} detected the planet using the direct imaging method and utilized RV data to calculate the orbital parameters of HD 206893 c. We try to mimic the signal identified by \citet{hinkley2023direct}, considering the RV data alone and using a jitter of 40 m/s appropriate for a young star. We note that the detected planet has an orbital period of 5.7 years which is longer than the 5.2 year coverage of the RV data. Notably, although there are 132 RV measurements, the majority are clustered within a few nights, providing effectively only 11 epochs to constrain the signal. Our fit provides a $\chi^2_r$ value of 4.09 and $\Delta$BIC value of 616.77 for the orbital solution of HD 206892 c which does not provide substantial support for its detection.\\
\\
We detect a periodic signal around HD 106906 with a period of 49.22$\pm$5.04 using TERRA RV data which is consistent with \citep{lagrange2016narrow}. We note that Gaia DR3 classifies HD 106906 as a single star based on its RUWE value of 0.93 \citep{vallenari2023gaia} but it should be noted that based on fig. 9 of \citet{Castro2024} that Gaia has little sensitivity to short-period near equal mass binaries. However, Lagrange et al. found that HD 106906 is an equal-mass spectroscopic binary system, consisting of two young F-type stars with nearly identical masses. The combined mass of these stars is approximately 2.6 $\si{M_{\odot}}$, and they orbit each other at a distance of 0.36$\pm$0.002 AU \citep{de2019near, rodet2017origin, lagrange2016narrow}. The TERRA pipeline assumes that HD 106906 is a single star (SB1), which provides one set of RV measurements.  However, HD 106906 is actually a spectroscopic binary system (SB2), e.g fig. 1 of \citep{de2019near} making it necessary to extract two sets of RVs$-$one for each star. Since this dataset would require a different data analysis approach to account for its binarity we consider this system as being beyond the scope of this paper. With additional radial velocity epochs as well as DR4 Gaia astrometry HD 1069060 should make a promising candidate for a full orbit fit \citep{gallenne2023araucaria}. We note that we also examined the TESS data and detected a consistent rotational period of 1.6 days across TESS sectors 11, 37, and 64. This period aligns with the periodic signal found by \citet{green202315}. \\
\\
Finally, six stars meeting our established criteria and displaying lower $RV_{RMS}$ values showed no RV-detected exoplanets. HD 48682, characterized by 92 RV data points—50 sourced from HIRES and 42 from HARPS; HIP 85235, represented by a collection of RV data points—17 from ELODIE and 66 from HIRES; HD 30495, featuring a compilation of 134 RV data points obtained from HARPS; HD 181327, showing 63 HARPS RV data points; HIP 29271, comprising an assemblage of 237 HARPS data points; and HD 53143, consisting of 90 HARPS RV data points.\\
\\
We now move forward to investigate the remaining eight stars. Among these, exoplanets have previously been detected using the RV method in six stars, while the remaining two stars have shown no exoplanet detections through the RV method. The stellar parameters employed to compute the attributes of potential exoplanets are outlined in Table \ref{stellar parameters}.

\begin{table*}
    \centering
    \begin{tabular}{ccccccccc}
        \hline
        Stars& Mass (\si{M_{\odot}})& Radius (\si{R_{\odot}}) & P$\rm_{Rot}$ (days)&[Fe/H] (dex)& Age (Gyr)&References\\
        \hline
        HD 10647 &$1.11\pm0.02$ & $1.10\pm0.02$ & $10\pm3$ &	$0.00\pm0.01$&	$1.4\pm0.9$& 1, 2, 15 \\
        HD 115617 &0.93 & $0.987\pm0.005$ & 29 &-0.02&	6.1 to 6.6&1, 3, 4, 16\\
        GJ 581 &$0.31\pm0.01$ & $0.310\pm0.008$& $132.5\pm6.3$ &$-0.33\pm0.12$ &	$8^{+3}_{-1}$&1, 5, 6, 7, 17\\
        HD 69830 &$0.89\pm0.03$ & $0.905\pm0.019$& $	35.1\pm0.8$ &	$-0.04\pm0.03$& $10.6\pm4.0$& 1, 8, 9, 10\\ 
        HD 142091 &$1.32\pm0.10$& $4.77\pm0.07$&$-$&	$0.13\pm0.03$&2.5&1, 20, 21 \\
        HD 207129 &$0.97_{-0.05}^{+0.07}$& 1.0& 12.6&$-$0.15 &	$3.8^{+3.6}_{-2.5}$&1, 11, 12, 13\\
        HD 202628 &$1.068\pm0.038$& $0.951\pm0.013$& $14.91\pm0.03$&	$0.00\pm0.06$&$2.3\pm1.0$&1, 14, 15, 18, 19 \\
        HD 22049 &0.82$\pm$0.02 & 0.735$\pm$0.005 &11.4 &$-$0.13$\pm$0.04&0.4 to 0.8& 22, 23, 24, 25, 26 \\
        \hline
    \end{tabular}
    \caption{Stellar parameters used to compute candidate exoplanet parameters around our target stars and also employed for statistical analysis. 1. \citep{vallenari2023gaia}, 2. \citep{ marmier2013coralie}, 3. \citep{perrin1988high}, 4. \citep{von2014stellar}, 5. \citep{ bean2006metallicities}, 6. \citep{pineda2021m}, 7. \citep{suarez_rotation_2017}, 8. \citep{ tanner2015stellar}, 9. \citep{laliotis2023doppler}, 10. \citep{ simpson2010rotation}, 11. \citep{marshall2011herschel}, 12. \citep{holmberg2007geneva}, 13. \citep{HD207129_CADARS}, 14. \citep{gaspar2016correlation}, 15. This Study, 16. \citep{mamajek2008improved}, 17. \citep{gliese581_ep_encyclopaedia}, 18. \citep{krist2012hubble}, 19. \citep{favata1997fe}, 20. \citep{white2018interferometric}, 21. \citep{bonsor2013spatially}, 22.\citep{benedict2006extrasolar}, 23. \citep{demory2009mass}, 24. \citep{roettenbacher2021expres}, 25. \citep{santos2004spectroscopic}, 26. \citep{van2007validation}.}
    \label{stellar parameters}
\end{table*}

\subsection{Analysis of Stars with Detected Exoplanets Using the Radial Velocity Method}

In this section, we present HD 10647, HD 115617, HD 69830, HD 142091, HD 22049, and GJ 581. Among these six target stars, we managed to employ an expanded dataset and revise the orbital parameters for four of them: HD 10647, HD 115617, HD 22049, and HD 142091.

\subsubsection{HD 10647}
\begin{figure}
\centering
\subfloat[] { \includegraphics[width=\linewidth]{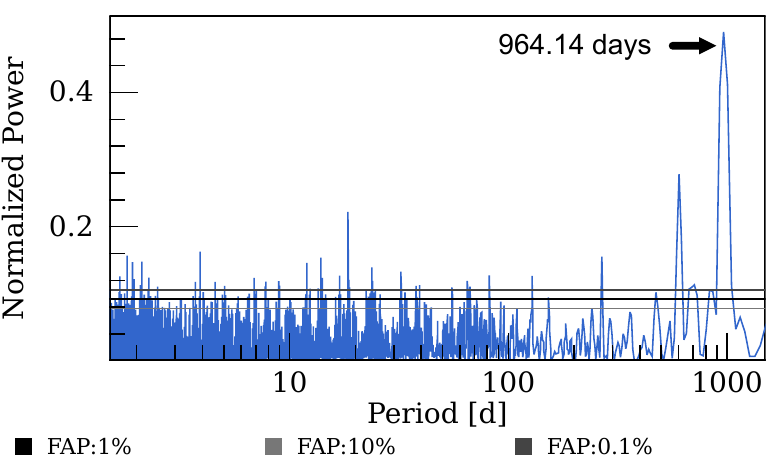} }
\\
\subfloat[] { \includegraphics[width=\linewidth]{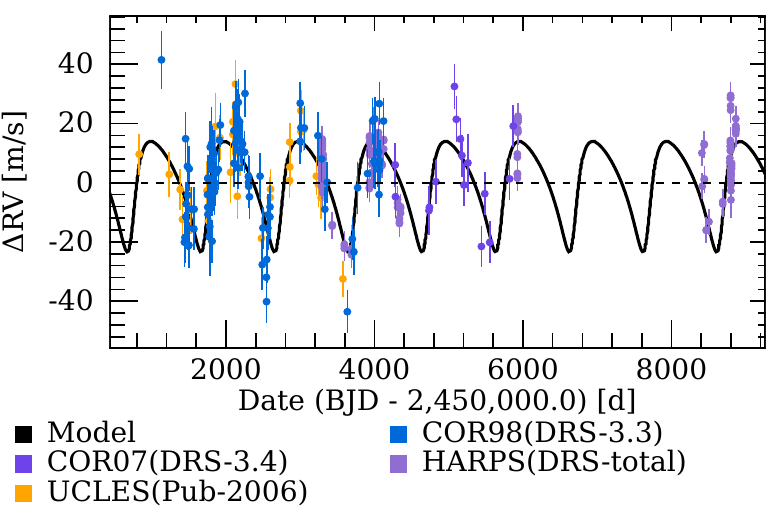}} 
\\
\subfloat[] { \includegraphics[width=\linewidth]{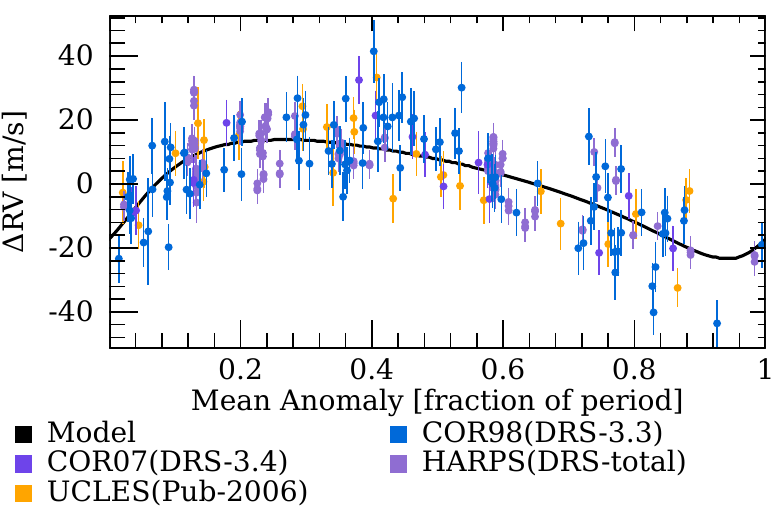}}
\caption{(a) LS periodogram of the RV data for HD 10647, exhibiting marked FAP values represented by three horizontal lines. Notably, a highly significant peak at 964.14 days is observed, signifying a periodic signal. (b) Doppler measurements of HD 10647 obtained from CORALIES, HARPS, and UCLES are presented as a scatter plot. The best-fit Keplerian model at $992.1\pm1.4$ days is superimposed on the data, illustrated by the black curve. (c) RV data and the one-planet Keplerian model phase-folded at 992.1 days. The phased data and our model demonstrate an excellent fit, confirming the presence of the detected periodic signal.}
\label{fig:model10647}
\end{figure}

HD 10647, classified as an F9V star \citep{lovell2021high}, exhibits a surrounding disc that was initially detected by the IRAS space-based telescope. The detection was based on the observation of an excess of infrared radiation \citep{stencel1991survey}. Extensive investigations by \citet{lovell2021high} and \citet{liseau2008q} have revealed that the disc possesses a high inclination, indicating an asymmetrical structure that extends predominantly in the northward direction. During the XIX IAP Colloquium Extrasolar exoplanets: Today and Tomorrow, an exoplanet named 10647b was first identified around HD 10647 by Michael Mayor in 2003. The existence of this planet was subsequently confirmed by \citet{butler2006catalog}. In 2013, \citet{marmier2013coralie} conducted a study on this planet, utilizing CORALIE RV data for their analysis.\\
\\
Our investigation commenced with an analysis of the stellar rotational period of HD 10647. \citet{mamajek2008improved} employed the stellar modulation technique and suggested a rotational period of $10 \pm 3$ days. In addition to this, we utilized TESS data to independently determine the rotational period. Our findings align with \citet{mamajek2008improved} so we adopt their value. For our RV analysis, we analysed 330 RV points acquired from multiple instruments, including HARPS, CORALIE98DRS-3.3 (COR98DRS-3.3), University College London Echelle Spectrograph (UCLES), and CORALIE07DRS-3.4 (COR07DRS-3.4). These RV measurements spanned the period from January 17, 1998, to January 17, 2020. We took into account the effect of RV jitter, \citet{butler2006catalog} calculated the RV jitter for HD 10647 to be 4.2 m/s. Additionally, we considered the error values in the RV measurements for each instrument. The LS periodogram for our RV data reveals a prominent peak at a period of 964.14 days, as depicted in Figure \ref{fig:model10647}a. We applied a one-Keplerian model to determine priors and perform MCMC to calculate the orbital paramters. Specifically, a single-planet Keplerian model was employed to analyze the RV dataset, as depicted in Figure \ref{fig:model10647}b, since no other notable peaks were observable. The optimal solution yielded a $\chi^2_r$= 2.31 and a $\Delta$BIC value of 1016. The $RV_{RMS}$ for the best fit was determined to be 7.84 m/s.\\
\\
In conclusion, we present a phase-folded Keplerian model with a period of $992.10\pm1.45$ days, visually confirming the observed periodic variations. Figure \ref{fig:model10647}c illustrates the phase-folded diagram. We employed a more extensive dataset of RV data points in comparison to \citet{marmier2013coralie} and \citet{butler2006catalog}, thereby facilitating the refinement of the orbital parameters. Additionally, we incorporated the inclination angle value of the debris disc around HD 10647 as mentioned in table \ref{main}, enabling the determination of $M_{DA}$ of HD 10647b. All orbital parameters for HD 10647b, calculated within this study and compared to those from the previous study by \citet{marmier2013coralie}, have been detailed in Table \ref{hd1067}.

\begin{table}
    \centering
    \begin{tabular}{ccc}
        \hline
        Parameter&\citet{marmier2013coralie} &This Study\\
        \hline
         P(days)&$989.2\pm8.1$ &991.90$\pm$1.45\\
         e&$0.15\pm0.08$ &0.37$\pm$0.03\\ 
         $\omega$(degree)& $212\pm39$&225.3$\pm$4.3\\
         K(m/s)&$18.1\pm1.7$ &18.44$\pm$0.91\\
         $T_P$(JD)&$2453654\pm99$& 2454700.68$\pm$10.82\\ 
         M$_{\rm P}$sin$i$ (M$_{\rm Jup}$)&$0.94\pm0.08$ &0.90$\pm$0.04\\
         $M_{DA}$ (M$_{\rm Jup}$)& &1.07$\pm$0.05\\
         Orbital Distance(AU) &$2.015\pm0.011$&2.02$\pm$0.01\\
         \hline
    \end{tabular}
    \caption{Orbital parameters of HD 10647b.}
    \label{hd1067}
\end{table}

\subsubsection{HD 115617}
The G6V star HD 115617 commonly referred to as 61 Vir, has garnered significant attention in recent years  \citep{gray2003contributions} . The Spitzer space telescope revealed the presence of a debris disc surrounding the star, manifesting as an infrared radiation excess at a wavelength of 160 $\mu$m \citep{wyatt2012herschel}. In December 2009, \citet{vogt2009super} provided the discovery of three exoplanets orbiting HD 115617, designated as 61 Vir b, 61 Vir c, and 61 Vir d. Initially, the existence of 61 Vir b and 61 Vir c was confirmed based on HARPS data in 2012 \citep{wyatt2012herschel}. However, at that time, the exoplanet signals associated with 61 Vir d remained unconfirmed. In 2021, \citet{rosenthal2021california} identified the signal as a yearly alias, suggesting it to be a false positive. Contrarily, a recent study conducted by \citet{laliotis2023doppler} proposed that the signals attributed to 61 Vir d do indeed represent exoplanet signals. \\
\\
\citet{baliunas1995chromospheric} reported an average rotation period of 29 days based on analysis of the Mt. Wilson survey data. We also generated a TESS LC for HD 115617 and then analysed it. However, the LC exhibited no discernible evidence of a periodic signal, offering no assistance in determining the rotational period of the star. Consequently, we use the \citet{baliunas1995chromospheric} value for the rotational period of the star. Our analysis incorporated a total of 3124 RV measurements obtained from various instruments, covering the period from April 30, 1991, to August 6, 2021. Specifically, we collected 179 RVs from APF, 171 RVs from HAMILTONPub-2014, 173 RVs from HAMILTONPub-2021, 602 RVs from HIRESPub-2021, 80 RVs from HIRESPub-2010, 520 RVs from HIRESPub-2017, 1273 RVs from HARPS, and  126 RVs from UCLESPub-2010. When considering the RV measurements, we accounted for the RV jitter. \citet{vogt2009super} showed a stellar jitter of 1.5 m/s for this star with the instrumental precision. Using these RV measurements, we generated LS periodograms for the merged RVs. To ensure the accuracy of our analysis, we disregarded any peaks at the stellar rotation period that could be attributed to stellar activity, focusing only on exoplanet signals. We identify three significant peaks corresponding to planetary systems. No other significant peaks were found apart from these three periods. A three-planet Keplerian model, aligned with the respective periods, was employed to ascertain the priors and perform MCMC. The goodness of our three-planet Keplerian MCMC model was calculated by a $\chi^2_r$ value of 1.44 and $\Delta$BIC of 8073.85. The $RV_{RMS}$ for the best fit was 2.56 m/s.\\
\\
We have compiled a comparative table detailing our observations, as outlined in Table \ref{61vira}. It is noteworthy that \citet{vogt2009super} employed 206 individual measurements from HIRES and AAT spanning a 16-year time series, while \citet{laliotis2023doppler} utilized a total of 1653 RV measurements from diverse instruments, including HIRES, UCLES, and HARPS, gathered between 2004 and 2020. In our investigation, we combined datasets from HIRES, HARPS, HAMILTON, and UCLES, encompassing a total of 3124 measurements taken between April 30, 1991, and August 6, 2021. Our outcomes demonstrate agreement with the prior findings, as all parameters fell within the uncertainty range reported by \citet{vogt2009super} and \citet{laliotis2023doppler}. However, due to the larger RV dataset at our disposal, we provide updated orbital parameters for these exoplanets. Additionally, by incorporating the inclination angle value from Table \ref{main}, we calculate $M_{DA}$ of the exoplanets.

\begin{table*}
    \centering
    \begin{tabular}{ccccc}
        \hline
        Planet & Parameter & \cite{vogt2009super}&\cite{laliotis2023doppler}& This Study\\
        \hline
         &P(days)& $4.2150\pm0.0006$ &$4.21498\pm0.00014$& 4.2150$\pm$0.0001\\
         &e& $0.12\pm0.11$ & $0.033\pm0.029$&0.11$\pm$0.03\\ 
         &$\omega$(degree)&  $105\pm54$ &&121.4$\pm$29.2\\
         61 Vir b&K(m/s)& $2.12\pm0.23$ & $2.47\pm0.11$&2.56$\pm$0.05\\
         &$T_P$(rjd)& 53369.166 &&55501.33$\pm$0.13\\
         &M$_{\rm P}$sin$i$ (M$_{\rm Earth}$)&	$5.1\pm0.5$ & $5.98^{+0.30}_{- 0.29}$  &6.11$\pm$0.24\\
         &$M_{DA}$ (M$_{\rm Jup}$)&&&0.197$\pm$0.001\\
         &Orbital Distance(AU)&	$0.050201\pm0.000005$ &&0.050$\pm$0.001\\
         \hline
         &P(days)& $38.021\pm0.034$ &$38.079\pm0.008$& 38.073$\pm$0.003\\
         &e& $0.14\pm0.06$& $0.026\pm0.023$& 0.07$\pm$0.01\\ 
         &$\omega$(degree)&$341\pm38$ && 327$\pm$14\\
         61 Vir c&K(m/s)& $3.62\pm0.23$ &$3.56\pm0.12$&3.87$\pm$0.06\\
         &$T_P$(rjd)& 53369.166 && 55487.45$\pm$1.47\\
         &M$_{\rm P}$sin$i$ (M$_{\rm Earth}$)&	$24.0\pm2.2$& $17.94\pm0.73$ & 19.33$\pm$0.70\\
         &$M_{DA}$ (M$_{\rm Jup}$)&   &&0.062$\pm$0.002\\
         &Orbital Distance(AU)&	$0.2175\pm0.0001$ &&0.216$\pm$0.004\\
         \hline
         &P(days)&  $123.01\pm0.55$ &$123.2\pm0.2$&123.12$\pm$0.08\\
         &e& $0.35 \pm 0.09$&$0.15\pm0.11$&0.12$\pm$0.03\\ 
         &$\omega$(degree)&$314\pm20$&&309$\pm$4\\
         61 Vir d&K(m/s)& $3.25\pm0.39$ & $1.47\pm0.17$&1.66$\pm$0.06\\
         &$T_P$(rjd)& 53369.166 &&55405.098$\pm$7.736\\
         &M$_{\rm P}$sin$i$ (M$_{\rm Earth}$)&	$18.2\pm1.1$  &$10.82^{+1.23}_{ -1.03}$&12.24$\pm$0.59\\
         &$M_{DA}$ (M$_{\rm Jup}$)& &&0.040$\pm$0.002\\
         &Orbital Distance(AU)&$0.476\pm0.001$& &0.47$\pm$0.01\\
         \hline
    \end{tabular}
    \caption{Orbital parameter of exoplanets detected around HD 115617.}
    \label{61vira}
\end{table*}

\subsubsection{HD 142091}

\citet{bonsor2013spatially} reported the detection of a debris disc surrounding HD 142091 ($\kappa$ Coronae Borealis). Additionally, \citet{johnson2008retired} identified an exoplanet orbiting this star using RV analysis. Subsequently, \citet{sato2012substellar} leveraged 96 RV data points obtained from the HAMILTON spectrograph \citep{spronck2013fiber} to corroborate the presence of this exoplanet. Finally, we integrate 416 RV points from HIRES, resulting in a compilation of 512 RV data points.\\
\\ 
For our observation, we utilized RV data spanning from April 20, 2004, to August 26, 2014. We commenced our observation by investigating the stellar jitter of this star, which had been reported as 2.84 m/s by \citet{luhn2020astrophysical}. Additionally, we took into account an instrumental precision and generated an LS periodogram for our RV data, revealing a significant peak at 1257.95 days. We also attempted to determine the rotational period of this star from existing literature and TESS data however no results were obtained. Given that we identified only one significant signal at 1257.95 days$-$unusually too long for a rotational period of a K-type star$-$ proceeded with our investigation. We employed a one-planet Keplerian model at the designated period and determine priors for the MCMC simulations. The fitting yielded a result, exhibiting a $\chi^2_r$ value of 0.69 and a $\Delta$BIC value of 3528.96. The $RV_{RMS}$ for the best fit was 2.81 m/s. Furthermore, we present our findings alongside an analysis conducted by \citet{baines2013navy}, who also incorporated the inclination angle of the disc to ascertain M$_{DA}$ of the identified planet. In line with their methodology, we provide an updated parameter assessment, utilizing an expanded dataset derived from earlier studies. The complete set of parameters is tabulated in Table \ref{hd142091}.

\begin{table}
    \centering
    \begin{tabular}{ccc}
        \hline
        Parameter&\citet{baines2013navy} &This Study\\
        \hline
         P(days)&$1300\pm15$ &1265.96$\pm$4.72\\
         e&$0.125\pm0.049$ &0.128$\pm$0.034\\ 
         $\omega$(degree)& &181.76$\pm$13.95\\
         K(m/s)&$27.3\pm1.3$ &25.74$\pm$0.77\\
         $T_P$(rjd)&&54819.83$\pm$589.60\\
         M$_{\rm P}$sin$i$ (M$_{\rm Jup}$)&$1.88\pm0.09$ &1.63$\pm$0.01\\
         $M_{DA}$ (M$_{\rm Jup}$)& 2.17 &1.92$\pm$0.01\\
         Orbital Distance(AU) &$2.8\pm0.1$&2.51$\pm$0.06\\
         \hline
    \end{tabular}
    \caption{Orbital parameters of HD 142091 b.}
    \label{hd142091}
\end{table}

\subsubsection{HD 22049}
HD 22049, also known as Epsilon Eridani, was identified by the Infrared Astronomical Satellite (IRAS) \citep{aumann1985iras} as having an infrared excess indicating the presence of circumstellar dust. Subsequent observations conducted with the James Clerk Maxwell Telescope (JCMT) at a wavelength of 850 $\mu$m unveiled an extended flux of radiation extending to an angular radius of 35 arcseconds around Epsilon Eridani, marking the first resolution of the debris disc. Subsequent higher-resolution imaging, employing ALMA, pinpointed the belt's location at 70 au from the star, with a width of merely 11 au \citep{ booth2023clumpy, booth2017northern}. Designated as Epsilon Eridani b, the existence of this planet was officially announced in 2000 \citep{hatzes2000evidence}, but its discovery stirred controversy over the ensuing two decades. In a 2008 study, the detection was labeled as "tentative," describing the proposed planet as "long suspected but still unconfirmed" \citep{backman2008epsilon}. Despite skepticism, numerous astronomers considered the evidence sufficiently compelling to confirm the discovery \citep{reidemeister2011cold, brogi2009dynamical, heinze2008deep}. However, doubts resurfaced in 2013 when a search program at La Silla Observatory failed to confirm the planet's existence \citep{zechmeister2013planet}. Subsequent investigations since 2018 gradually reaffirmed the planet's presence through a combination of RV and astrometry \citep{llop2021constraining, makarov2021looking}. In our analysis, we utilized a total of 2192 RV data points from instruments such as HARPS, HIRES, HAMILTON, and APF, and during this examination, we successfully detected Epsilon Eridani b.\\
\\
We analyzed RV data spanning from September 1987 to February 2020. As the available literature lacked information on the stellar jitter, we computed the RV jitter for this star following the methodology outlined in \citet{isaacson2010chromospheric}. We utilized values from \citet{jenkins2006activity}, resulting in a calculated stellar jitter of 3.56 m/s. Additionally, we considered instrumental precision and generated an LS periodogram for our RV data, revealing a highly significant peak at 2774.41 days. Furthermore, we took into account the rotational period of the star, determined to be 11.4 days based on findings by \citet{roettenbacher2021expres}. To further investigate the potential exoplanet signal, we employed a one-planet Keplerian model at the identified period and determined priors for MCMC simulation. The fitting process resulted in an outcome, with a $\chi^2_r$ value of 2.80 and a $\Delta$BIC value of 4870.1. $RV_{RMS}$ for the best fit was 7.59 m/s. Moreover, we present our findings alongside an analysis conducted by \citet{feng2023revised}, who used orbital inclination value derived using astrometric data and calculated the mass of the planet. A detailed summary of all parameters is presented in Table \ref{hd22409}.

\begin{table}
    \centering
    \begin{tabular}{ccc}
        \hline
        Parameter&\citet{feng2023revised} &This Study\\
        \hline
         P(days)&$2688.60^{+16.17}_{-16.51}$ &2806.04$\pm$5.55\\
         e&$0.26\pm0.04$ &0.16$\pm$0.01\\ 
         $\omega$(degree)& $224.37\pm 5.87$&225.3$\pm$4.3\\
         K(m/s)&$9.98\pm0.43$ &11.37$\pm$0.20\\
         $T_P$(RJD)&$44411.54\pm76.60$& 52785.68$\pm$61.09\\ 
         M$_{\rm P}$sin$i$ (M$_{\rm Jup}$)&& 0.68$\pm$0.03\\
         M$_{\rm P}$ (M$_{\rm Jup}$)& $0.76^{+0.14}_{-0.11}$&\\
         $M_{DA}$ (M$_{\rm Jup}$)&&0.70$\pm$0.03\\
         Orbital Distance(AU) &$3.53\pm0.06$&3.64$\pm$0.06\\
         \hline
    \end{tabular}
    \caption{Orbital parameters of HD 22049b.}
    \label{hd22409}
\end{table}

\subsubsection{HD 69830}

HD 69830, a G8V star, has been the subject of significant astronomical investigation. The presence of a narrow ring of warm debris encircling the star was initially detected by the Spitzer Space Telescope in 2005 \citep{beichman2005excess}. Subsequent exploration led to the confirmation of three exoplanets with minimum masses similar to those of Neptune. These exoplanets are all situated within the orbit of the debris ring.  In our study, we utilized 648 RVs from the HARPS instrument and 439 RVs from HIRES, resulting in a total of 1087 RVs data points. However, \citet{laliotis2023doppler} employed a more extensive dataset, consisting of 1589 RV data points, which represents an update to our analysis. We also utilize the disc inclination angle to derive the $M_{DA}$ of the planet. Additionally, we determined the orbital distance using the dataset available to us, and we have summarized the relevant parameters in Table \ref{HD69830table}.

\begin{table}
    \centering
    \begin{tabular}{cccc}
        \hline
        Planet & Parameter &Best fit& Ref\\
        \hline
         &P(days)&$8.66897\pm0.00028$&1\\
         &e& $0.128\pm0.028$&1\\ 
         HD 69830 b &K(m/s) & $3.4\pm0.1$ &1\\
         &M$_{\rm P}$sin$i$ (M$_{\rm Earth}$)&$10.1^{+0.38}_{-0.37}$ &1\\
         &$M_{DA}$ (M$_{\rm Jup}$)&$0.14^{+1.75}_{-0.34}$&2\\
         &Orbital Distance(AU)&$0.079\pm0.001$ &2\\
         \hline
         &P(days) &$31.6158\pm0.0051 $&1\\
         &e& $0.03\pm 0.03$&1\\ 
         HD 69830 c&K(m/s)&$2.6\pm0.1$&1\\
         &M$_{\rm P}$sin$i$ (M$_{\rm Earth}$)& $12.09^{+0.55}_{-0.54}$&1\\
         &$M_{DA}$ (M$_{\rm Jup}$)&  $0.17^{+2.12}_{-0.11}$ &2\\
         &Orbital Distance(AU)& $0.188\pm0.001$&2\\
         \hline
         &P(days)&$201.4\pm0.4$&1\\
         &e&$0.08\pm0.07$&1\\ 
         HD 69830 d&K(m/s)&$1.5\pm 0.1$&1\\
         &M$_{\rm P}$sin$i$ (M$_{\rm Earth}$)&$12.26^{+0.89}_{-0.88}$&1\\
         &$M_{DA}$ (M$_{\rm Jup}$)&$0.17^{+2.17}_{-0.11}$ &2\\
         &Orbital Distance(AU)&$0.644\pm0.001$&2\\
         \hline
    \end{tabular}
    \caption{Orbital parameter of exoplanets detected around HD 69830. The source of the values in the table (Ref) is given in the final column:
    1.\citep{laliotis2023doppler}, 2. This study.}
    \label{HD69830table}
\end{table}

\subsubsection{GJ 581}
GJ 581 is categorized as an M3V star with a mass roughly one-third that of the Sun \citep{reyle202110}. The presence of a debris disc around GJ 581 was noted within the DEBRIS program conducted on the Herschel Space Observatory \citep{lestrade2012debris}. The initial planet in orbit around GJ 581 was discovered in August 2005 \citep{bonfils2005harps}. Following this discovery, the same group identified two additional exoplanets, Gliese 581c and Gliese 581d \citep{udry2007harps}. Another planet, Gliese 581e, was announced on April 21, 2009, \citep{mayor2009harps}. Subsequent findings introduced Gliese 581g and Gliese 581f as two more exoplanets orbiting this star \citep{vogt2010lick}. Although later investigations indicated that Gliese 581f was more likely associated with the stellar activity cycle \citep{trifonov2018carmenes, robertson2013halpha, ballard2013exoplanet} argued for the existence of three exoplanets encircling GJ 581$-$specifically Gliese 581b, Gliese 581c, and Gliese 581e. Our analysis also confirmed the presence of three exoplanet signals, consistent with the discoveries of \citet{trifonov2018carmenes}. Given the absence of updated RV data, we reference \citet{trifonov2018carmenes} in Table \ref{GJ581table}. Furthermore, we incorporated the debris disc inclination angle value from Table \ref{main} and determined $M_{DA}$ of the detected exoplanets.
\begin{table}
    \centering
    \begin{tabular}{cccc}
        \hline
        Planet & Parameter &Best fit&Ref\\
        \hline
         &P(days)&$5.368\pm0.001$&1\\
         &e&$0.022^{+0.027}_{-0.005}$&1\\ 
         &$\omega$(degree)&  $118.3^{+27.4}_{-22.9}$&1\\
         Gliese 581 b&K(m/s)&$12.35^{+0.18}_{-0.20}$ &1\\
         &M(deg)&$163.4^{+22.9}_{-23.9}$&1\\
         &M$_{\rm P}$sin$i$ (M$_{\rm Earth}$)&$15.20^{+0.22}_{-0.27}$&1\\
         &$M_{DA}$ (M$_{\rm Jup}$)&$0.048\pm0.001$&2\\
         &Orbital Distance(AU)& $0.041\pm0.001$&1\\
         \hline
         &P(days)&$12.919^{+0.003}_{-0.002}$&1\\
         &e&$0.087^{+0.150}_{-0.016}$&1\\ 
         &$\omega$(degree)&$148.7^{+71.5}_{-33.0}$&1\\
         Gliese 581 c&K(m/s)&$3.28^{+0.22}_{-0.12}$&1\\
         &M(deg)&$218.0^{+37.3}_{-68.4}$&1\\
         &M$_{\rm P}$sin$i$ (M$_{\rm Earth}$)&$5.652^{+0.386}_{-0.239}$&1\\
         &$M_{DA}$ (M$_{\rm Jup}$)&$0.018\pm0.001$&2\\
         &Orbital Distance(AU)& $0.074\pm0.001$&1\\
         \hline
         &P(days)&$3.153^{+0.001}_{-0.006}$&1\\
         &e&$0.125^{+0.078}_{-0.015}$&1\\ 
         &$\omega$(degree)&$77.4^{+23.0}_{-43.6}$&1\\
         Gliese 581 e&K(m/s)&$1.55^{+0.22}_{-0.13}$&1\\
         &M(deg)&$203.7^{+56.6}_{-21.4}$&1\\
         &M$_{\rm P}$sin$i$ (M$_{\rm Earth}$) &$1.657^{+0.240}_{-0.161}$&1\\
         &$M_{DA}$ (M$_{\rm Jup}$) &$0.005\pm0.001$&2\\
         &Orbital Distance(AU)& $0.029\pm0.001$&1\\
         \hline
    \end{tabular}
    \caption{Orbital Parameters of exoplanets Detected around GJ 581. 1.\citep{trifonov2018carmenes}, 2. This study.}
    \label{GJ581table}
\end{table}

\subsection{Analysis of stars without detected exoplanets using the Radial Velocity Method}
In the subsequent sections, we provide a detailed analysis of two of these stars, namely HD 207129 and HD 202628, elucidating their RV characteristics and potential for hosting undiscovered companions.

\subsubsection{HD 207129}
\begin{figure}
    \centering
    \subfloat[] { \includegraphics[width=\linewidth]{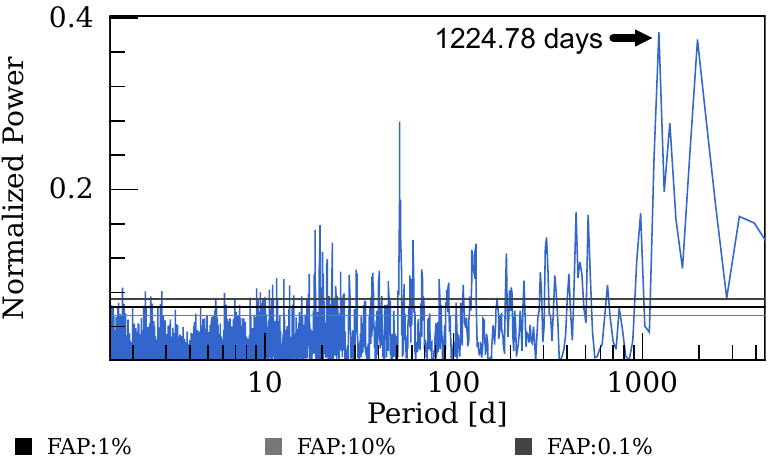}}
    \\
    \subfloat[ ] { \includegraphics[width=\linewidth]{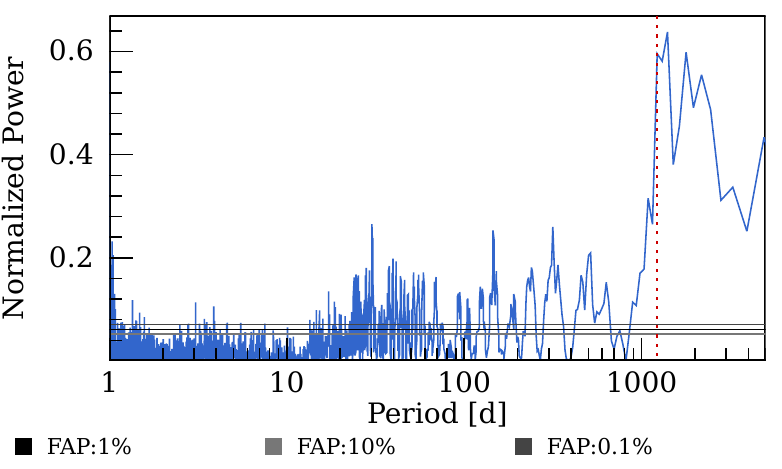}}
    \\
    \subfloat[ ] { \includegraphics[width=\linewidth]{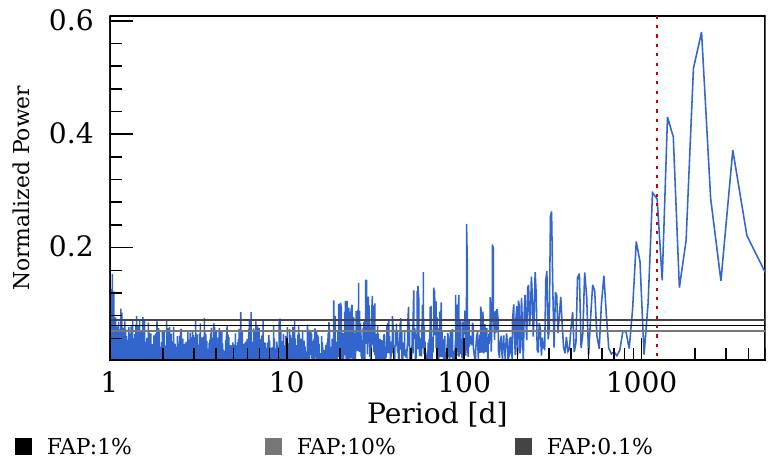}}
    \caption{Panel a: LS Periodogram of HARPS RV data for HD 207129, emphasizing the most prominent peak at 1224.78 days (indicated by an arrow). Panel b: LS Periodogram for the S-Index Indicator of HD 207129, with the dotted red line representing the 1224.78 days period. Panel c: LS Periodogram for the BIS Indicator, with the dotted red line indicating the 1224.78 days period.}
    \label{fig:pf207129}
\end{figure}

HD 207129 is a G2V star and has been the subject of study using the Advanced Camera for Surveys (ACS) instrument on the Hubble Space Telescope. Notably, imaging of a debris disc around HD 207129 has been accomplished both in visible light using the ACS instrument and in the infrared (70 $\mu m$) using the MIPS instrument on the Spitzer Space Telescope \citep{krist2010hst}.\\
\\
Before our investigation, no exoplanets had been detected in close proximity to HD 207129. To conduct our analysis, we incorporated 482 RV points from HARPS with a stellar jitter value of 1.76 m/s \citep{hojjatpanah2020correlation}. Considering the instrumental precision, we focused on determining the rotational period of HD 207129, which was found to be 12.6 days \citep{marshall2011herschel}. Additionally, we utilized TESS data to search for the rotational period but did not find any periodic signal, so we proceeded with the rotational period value from the literature for our analysis. In our investigation of periodic signals in the RV data, we identified several significant peaks with FAP better than $0.1\%$ as shown in figure \ref{fig:pf207129}a. To explore potential exoplanet signals further, we performed a Keplerian fit to the three most significant peaks observed at periods of 51.57, 1224.78, and 1959.65 days. A single-planet Keplerian model was applied to all three peaks, but we observed highly eccentric orbits (e $\geq$ 0.5), except for the most significant peak at 1224.78 days, leading us to consider it a candidate planetary peak. In order to confirm whether the observed signal indicates a planetary presence or merely activity, we conducted an analysis using the S-Index and BIS indicators. Subsequently, we plotted LS periodograms for both indicators, as depicted in Figure \ref{fig:pf207129}b and Figure \ref{fig:pf207129}c, respectively. The periodograms for both the S-Index and BIS indicators reveal prominent peaks occurring at 1224.78 days. The normalized powers associated with these peaks are 0.6 for the S-Index and 0.29 for BIS. Upon comparing these power values with the normalized power of the LS periodogram in Figure \ref{fig:pf207129}a, which stands at 0.38, it becomes evident that the S-Index activity signal significantly surpasses the RV signal. Therefore, based on this comparison, we conclude that the observed signal is more likely attributable to a possible activity rather than an exoplanet signal.

\subsubsection{HD 202628}
\begin{figure}
    \centering
    \subfloat[] { \includegraphics[width=\linewidth]{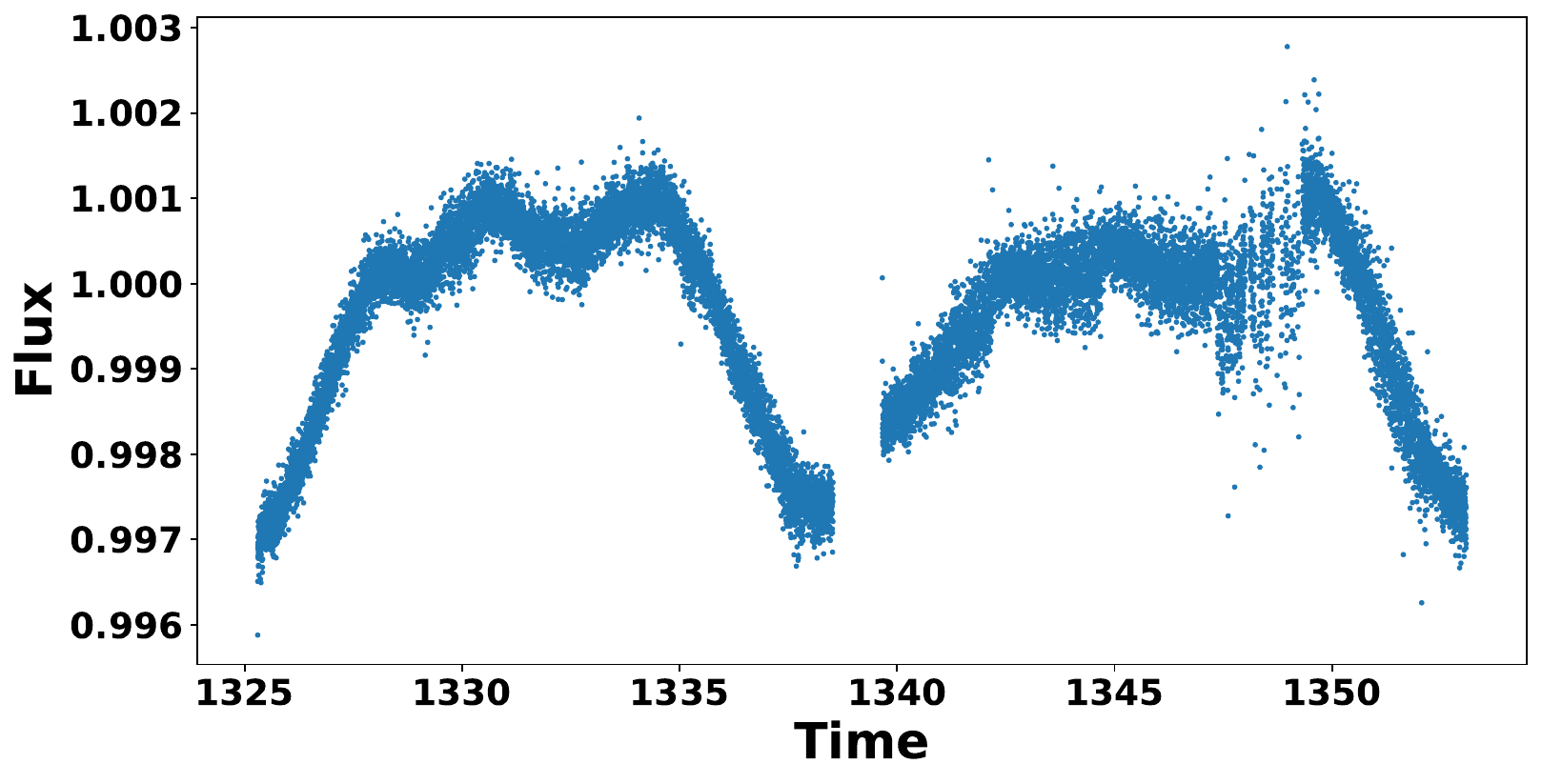}}
    \\
    \subfloat[ ] { \includegraphics[width=\linewidth]{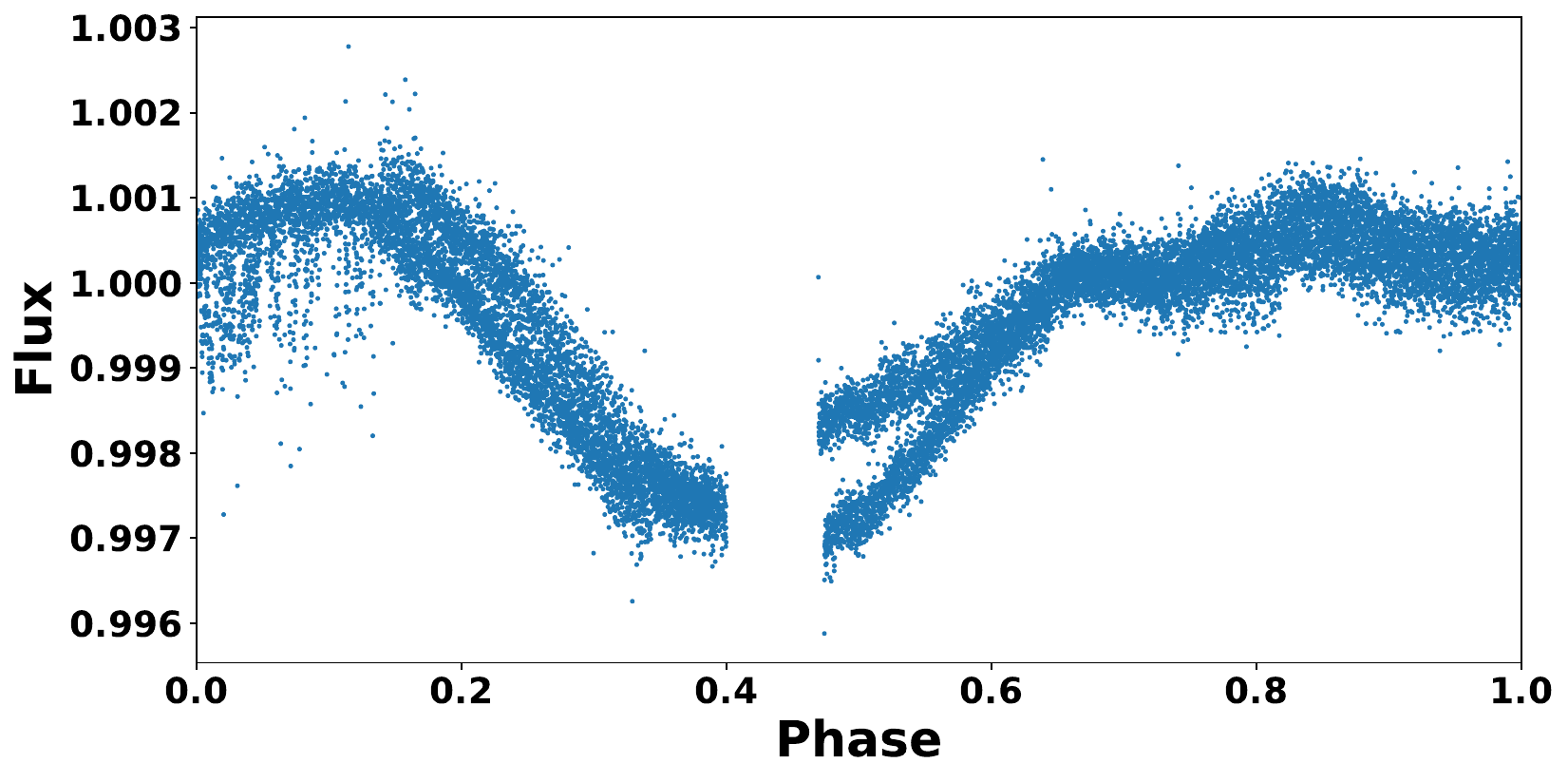}}
    \\
    \subfloat[ ] { \includegraphics[width=\linewidth]{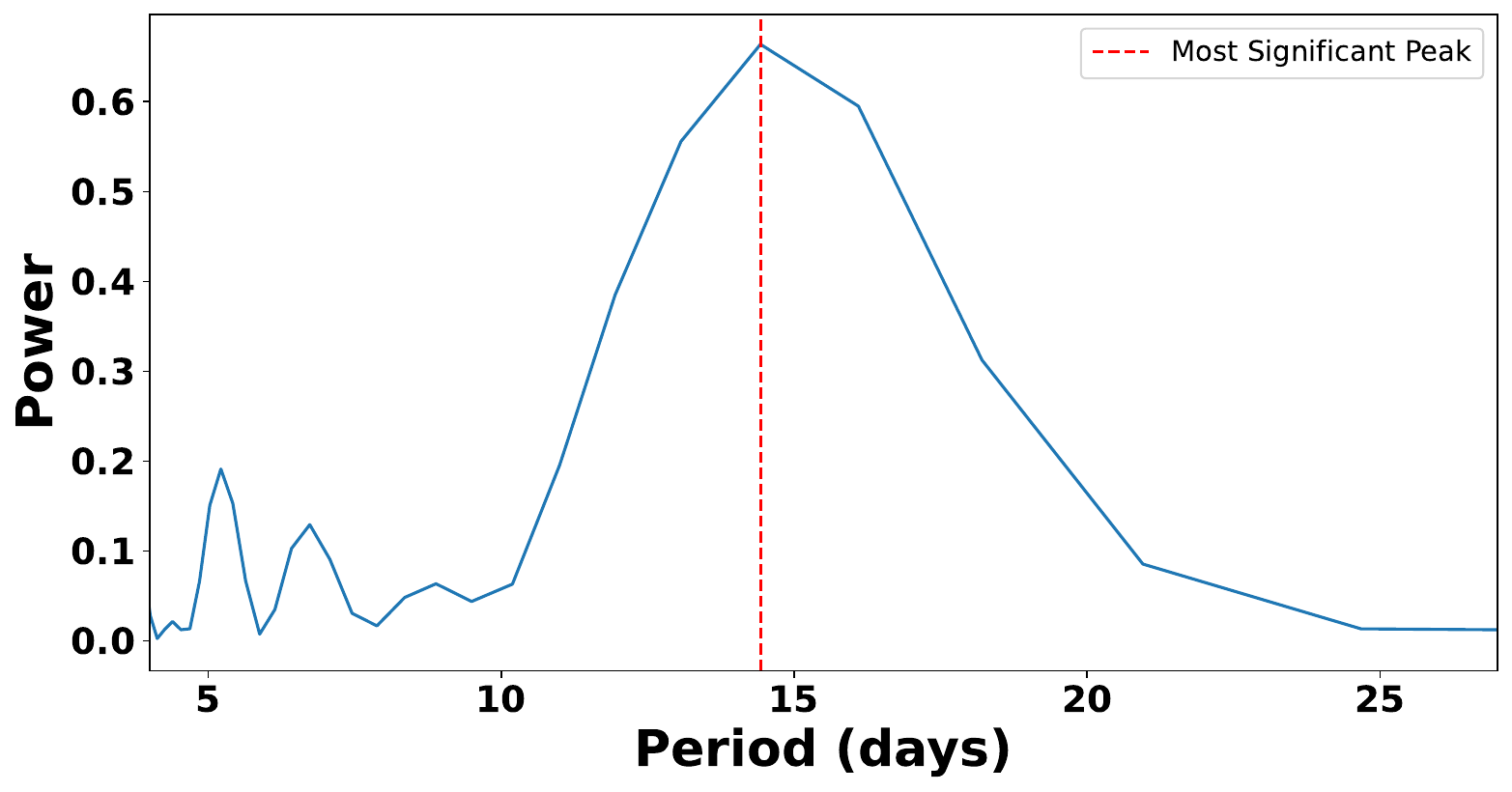}}
    \caption{Panel a displays the TESS LC of HD 202628, illustrating the temporal variation in the star's flux as observed by TESS. Panel b exhibits the flux phase folding of HD 202628 over a period of 14.91 days. The Flux phase folding plot demonstrates a dip in brightness, indicating a significant variation occurring at this specific period. Panel c shows the LS periodogram derived from the TESS LC of HD 202628. The LS periodogram reveals the spectral power distribution of periodic signals present in the LC data. The most significant peak in this periodogram analysis is represented by the red dotted line at a period of $14.91\pm0.03$ days.}
    \label{fig:TESSmodel202628}
\end{figure}

\begin{figure}
    \centering
    \subfloat[] { \includegraphics[width=\linewidth]{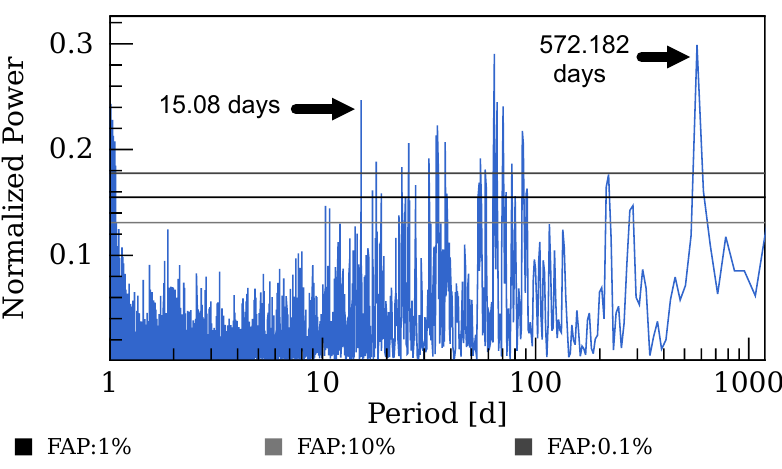}}
    \\
    \subfloat[ ] { \includegraphics[width=\linewidth]{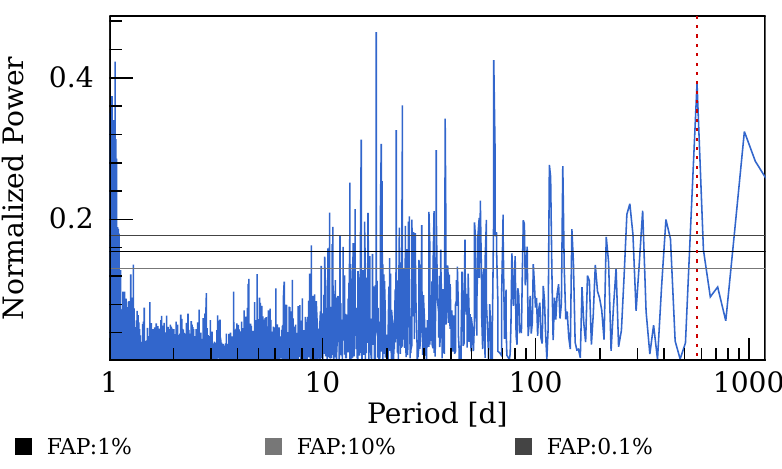}}
    \\
    \subfloat[ ] { \includegraphics[width=\linewidth]{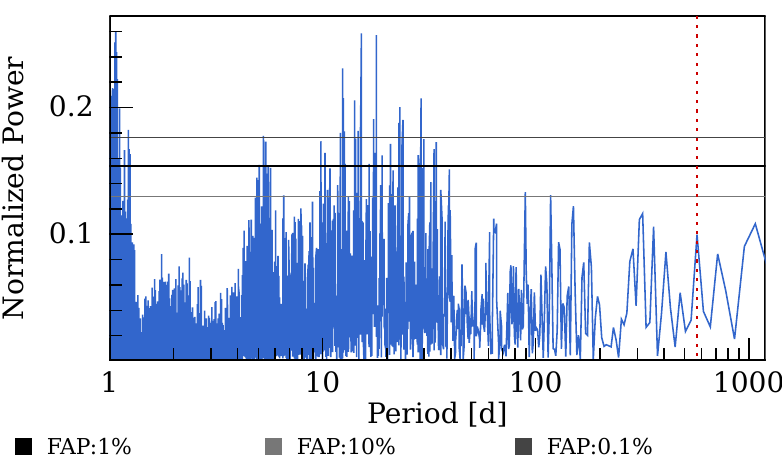}}
    \caption{Panel a: LS periodogram of the HARPS RV dataset for HD 202628. The peak at 15.08 days corresponds to the rotational period, while the peak at 572.18 days indicates the presence of a potential exoplanet signal. Panel b: LS Periodogram for S-Index indicator where dotted red line denotes the 572.18 days signal. Panel c: LS Periodogram for the BIS Indicator, with the 572.18-day period marked by the dotted red line}
    \label{fig:model202628}
\end{figure}

HD 202628 is a G2V star similar to that of the Sun \citep{faramaz2019scattered}. In 2010, the Spitzer space telescope detected infrared excess from a circumstellar disc around this star. On analysing the disc, \citet{faramaz2019scattered} saw a sharply defined inner edge of the disc at a distance of 150 AU, and \citet{nesvold2014gap} showed that an exoplanet is responsible for this sharp inner edge. \\
\\
No literature regarding the rotational period of the star HD 202628 was found. We utilized data from TESS to determine the stellar rotational period of HD 202628. For our analysis, we extracted TESS data using the SPOC pipeline, as suggested by \citet{martins2020search}. The light curves (LCs) of each TESS sector were then de-trended with third-order polynomial fits. This step serves as a high-pass filter, suppressing long-term trends typically associated with instrumental systematics. Subsequently, we performed the removal of outliers by excluding any flux measurement greater than 3.5 times the standard deviation of the de-trended LCs. Upon extracting the TESS data, observations were available in two sectors, namely Sector 1 and Sector 29. Our intention was to combine these two sectors to form a long-term series. However, inspected of the individual sectors shows the variation observed in sector 1, as shown in Figure \ref{fig:TESSmodel202628}a, was absent in sector 29. In sector 29, the light curve appeared almost linear. The reason for this discrepancy is attributed to the absence of starspots, not unexpected given the approximately two-year time difference between the observations in different sectors puts a Sun-like star in a different phase of its activity cycle. Consequently, for our analysis, we focus on TESS Sector 1 data. Here a rotational period of 14.91 days was detected, accompanied by a low FAP of less than 0.01$\%$. Subsequently, we phase-folded the dataset using the identified period of 14.91 days, as demonstrated in Figure \ref{fig:TESSmodel202628}b. This phase-folded representation exhibited a distinct dip, providing evidence for a stellar rotational period. We employed an LS periodogram on the LC data, depicted in Figure \ref{fig:TESSmodel202628}c. The LS periodogram indicated a peak at a period of $14.91\pm0.03$ days, which we identify as the rotational period of the star.\\
\\
Our RV analysis involved examining 179 measurements obtained from HARPS. As the literature lacked information regarding the stellar jitter, we calculated the RV jitter for this star following the methodology mentioned in \citet{isaacson2010chromospheric} and utilizing the S$_{HK}$ values from \citet{jenkins2006activity}, resulting in a stellar jitter of 4.16 m/s. We also considered instrumental errors, to validate the detected stellar rotational period of $14.91\pm0.03$ days, we constructed an LS periodogram for the RV data, revealing a significant peak with a FAP better than 0.1 percent at a period of 15.08 days, as depicted in Figure \ref{fig:model202628}a. This finding reaffirmed the accuracy of the identified stellar rotational period. We further explored potential exoplanet signals, observing a notable peak at 572.182 days in Figure \ref{fig:model202628}a. Subsequently, we sought the presence of the 572.182-day signal in both the S-Index and BIS indicators. We identify a signal in the S-index, as illustrated by dotted red line in Figure \ref{fig:model202628}b though no signal is evidence in the BIS index \ref{fig:model202628}c. Conservatively, we interpret this signal as indicative of possible activity rather than an exoplanet signal.

\section{Discussion}
\begin{figure*}
    \centering
    \subfloat[] { \includegraphics[width=0.5\textwidth]{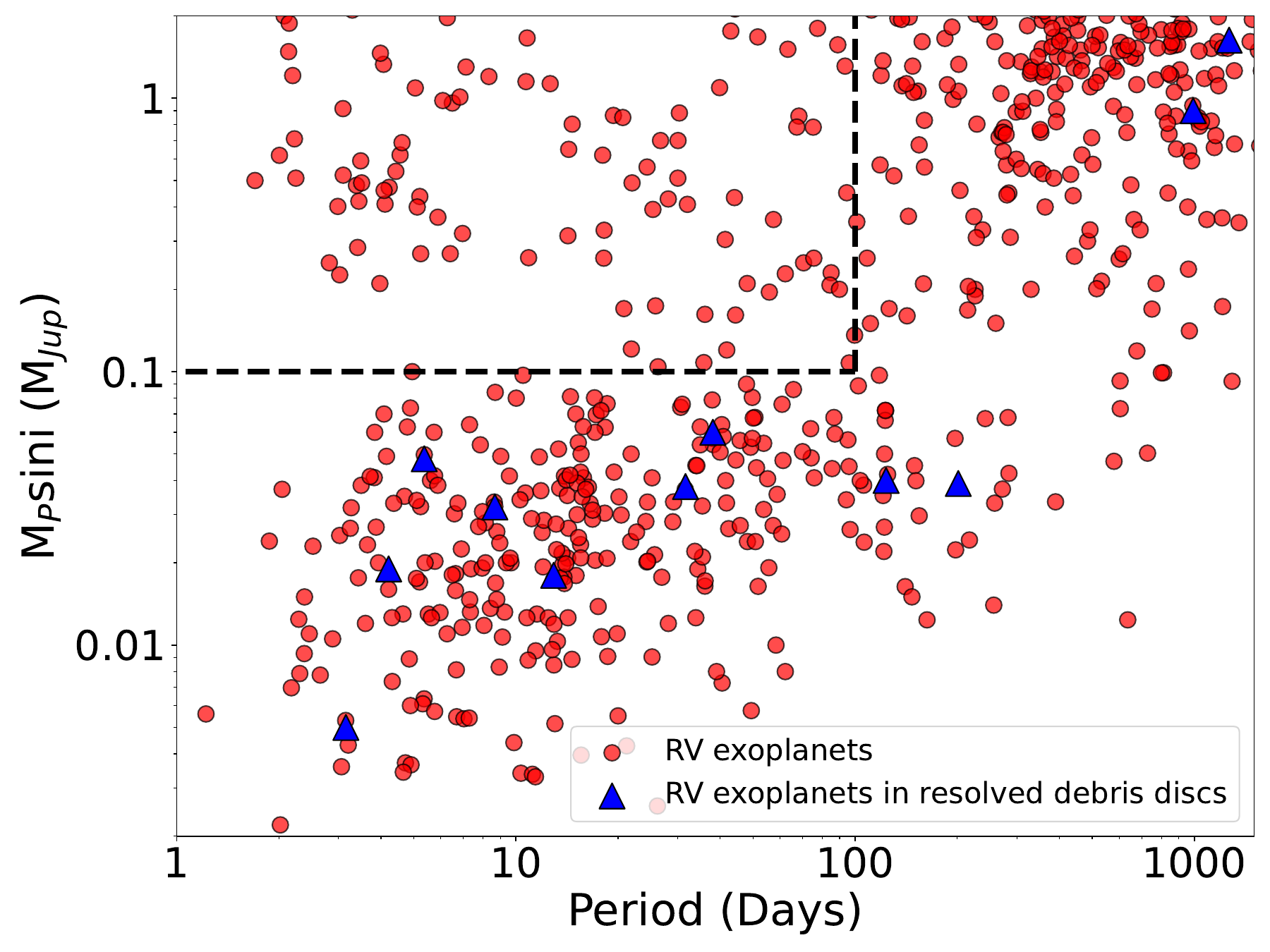}}
    \subfloat[]{ \includegraphics[width=0.5\textwidth]{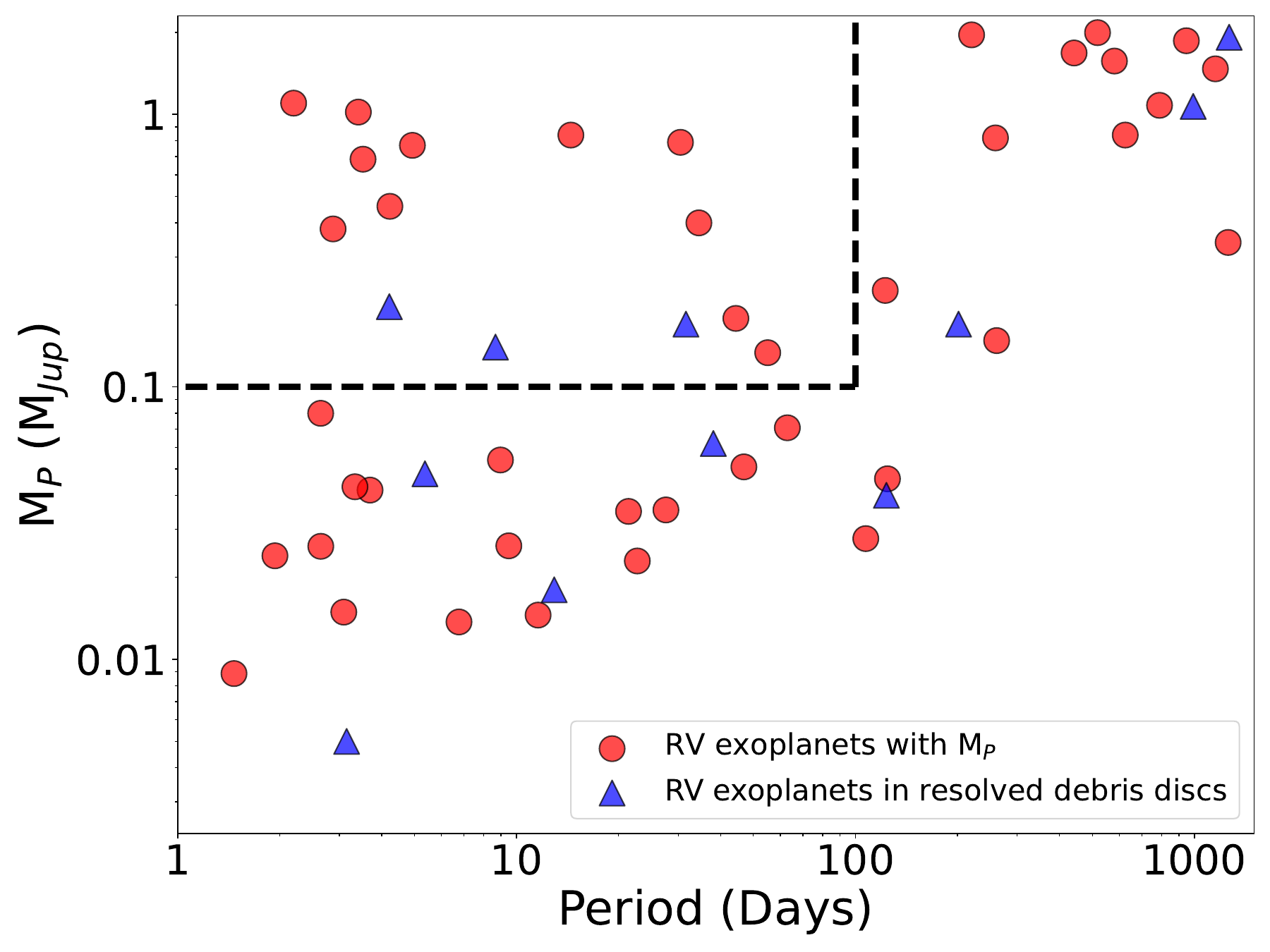}}\\
    \subfloat[]{ \includegraphics[width=0.5\textwidth]{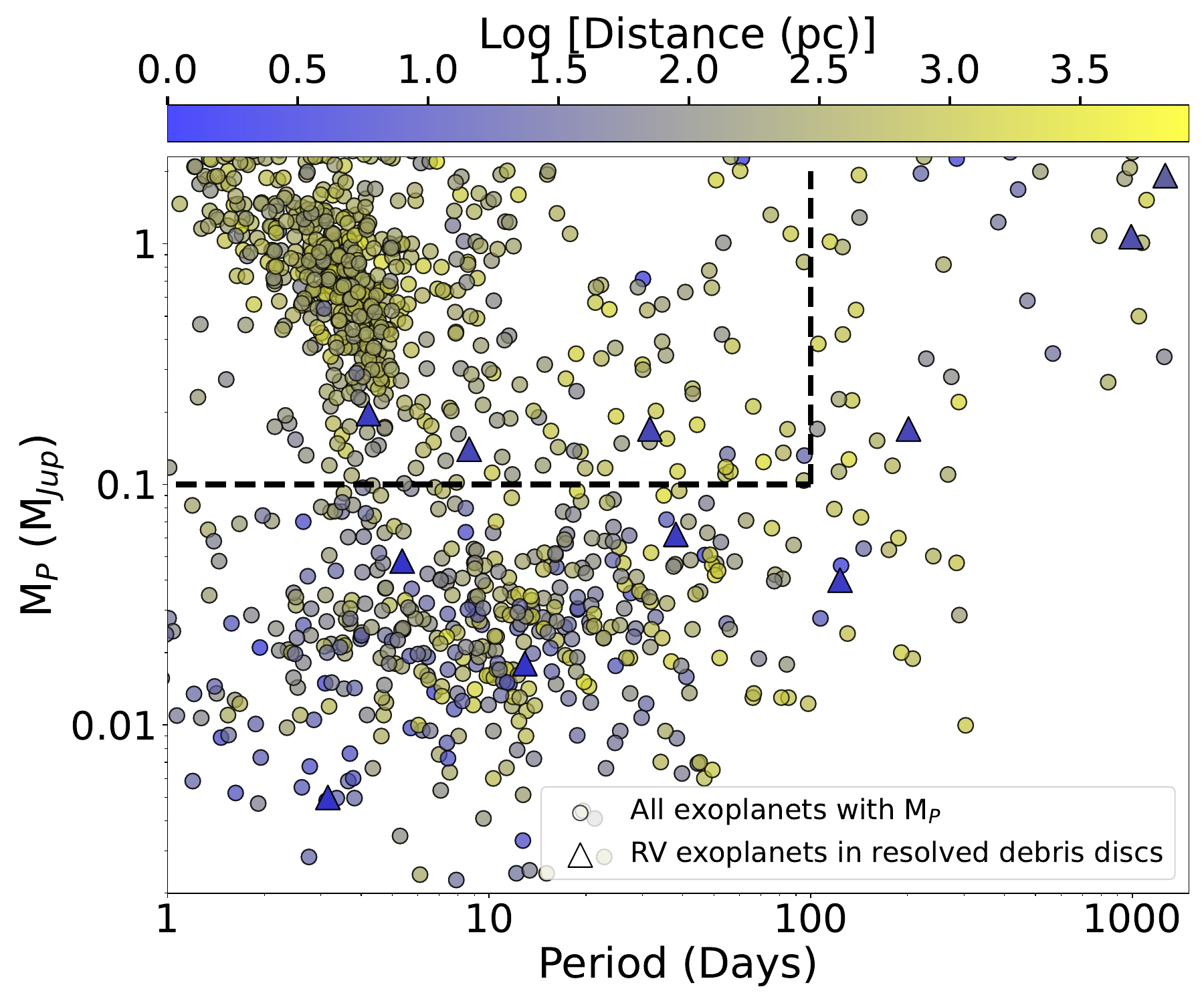}}
    \caption{Panel a: Scatter plot illustrating the correlation between M$_{\text{P}}$ $\sin{i}$ and the orbital period. Red circles represent RV-detected exoplanets sourced from the NASA Exoplanet Archive \citep{akeson2013nasa}, while blue triangles denote exoplanets detected within resolved debris discs as part of this study. The region marked by dotted lines indicates the area for warm and hot Jupiters. Panel b: Scatter plot depicting the correlation between M$_{\text{P}}$ and the orbital period, focusing exclusively on RV-detected exoplanets. Red dots represent true masses sourced from the NASA Exoplanet Archive, while blue triangles signify M$_{DA}$ of the RV-detected exoplanets discovered within the scope of this study. The region marked by dotted lines denotes the area for warm and hot Jupiters. Panel c: Scatter plot to depict the correlation betwee  \let\Bbbk\relax n M$_{\text{P}}$ and the orbital period. Circles represent exoplanets for which true masses have been previously detected, sourced from the NASA Exoplanet Archive. Triangles depict $M_{DA}$ of the RV-detected exoplanets within resolved debris discs as part of this study. The color coding on this plot corresponds to the log of the stellar distance of the respective star around which the corresponding planet has been detected. }
    \label{fig:statanalysis}
\end{figure*}

From our selection of 29 stars, we achieved successful validation of exoplanet signals for HD 10647, HD 115617, HD 69830, HD 142091, HD 22049, and GJ 581. Additionally, we refined the orbital parameters for exoplanets orbiting four specific stars$-$HD 10647, HD 115617, HD 22049, and HD 142091$-$ by employing a more extensive dataset. Furthermore, through the utilization of the debris disc inclination angle for each star, we calculated $M_{DA}$ of the exoplanets orbiting HD 10647, HD 115617, HD 69830, GJ 581, HD 22049, and HD 142091. We have also detected long-term activity signals in HD 207129 and HD 202628. Finally, we compared our findings with prior studies. One of our assumptions was to consider disc-planet coplanarity and determine M$_{DA}$. Upon examining the results of HD 22049, we noticed that \citet{feng2023revised} utilized astrometric data to compute the orbital inclination and calculate the mass of a planet. When comparing this mass value to the M$_{DA}$ value of the planet detected around HD 22049, we find that the M$_{DA}$ value of 0.70$\pm$0.03 M$_{\rm Jup}$ closely aligns with the mass value of 0.76$^{+0.14}_{-0.11}$ M$_{\rm Jup}$, supporting our assumption of disc-planet coplanarity for estimating the closest mass approximation to the true mass of a planet.  Furthermore, we observe that none of our young A$-$type stars exhibit any evidence of exoplanets when compared to solar-type stars. However, this absence may be attributed to detection bias; all of these stars have rapid rotation, which reduces radial velocity precision, as indicated by their high values of $RV_{\rm RMS}$ in Table \ref{main}.\\
\\
Within our dataset,  five out of six stars with exoplanets harbor at least one low-mass planet with a minimum mass (M$_{\rm P}$sin$i$) of $\leq$ 30 M$_{\rm Earth}$, and four out of these five stars have a metallicity ([Fe/H]) $\leq$ 0. This metallicity trend is consistent with the trends observed by \citet{maldonado2012metallicity, sousa2011spectroscopic}, which suggests that exoplanets hosting low-mass exoplanets tend to have a metallicity less than solar. Referring to the predictions of \citet{fernandes2019hints}, the occurrence rate of exoplanets with masses (M$_{\rm P}$sin$i$) ranging from 0.1 to 20 M$_{\rm Jup}$ and orbital distances between 0.1 and 100 AU was $26.6\%_{-5.4\%}^{+7.5\%}$. In our analysis, we have identified three exoplanets that fall within this range. Similarly, \citet{mayor2009harpsss} suggested that the occurrence rate of super-Earths and Neptunes (with M$_{\rm P}$sin$i$=3-30 M$_{\rm Earth}$) having orbital periods of less than 50 days is estimated to be $30\% \pm 10\%$. Within our dataset, five exoplanets fall within this specified range.\\
\\
\citet{maldonado2012metallicity} suggest that out of 29 planet-hosting stars with debris discs, 11 exhibit multiplanet systems, resulting in an incidence rate of 38$\%$. Similarly, \citet{cao2023catalog} report an incidence rate of 40$\%$, where out of 73 discs with exoplanets, 29 host multiplanet systems. In our analysis, out of the seven planet-hosting stars we find with resolved debris discs, three of them host multiplanet systems. Additionally, \citet{wolthoff2022precise} determined that the completeness-corrected global occurrence rate of giant planetary systems (systems with at least one planet with  M$_{\rm P}$sin$i$ $>$ 0.8 M$_{\rm Jup}$) around evolved stars is $10.7\%^{+2.2\%}_{-1.6\%}$. In comparison, within our dataset of seven planetary systems, two exoplanets (HD 10647b, HD 142091 b) have  M$_{\rm P}$sin$i$ greater than 0.8 M$_{\rm Jup}$. The limited size of our dataset constrains our ability to make strong conclusions about the exoplanet population surrounding debris discs. Overall, we do not find any significant indications of differences relative to earlier findings.\\
\\ 
For visual analysis, we generated three scatter plots. The exoplanets selected for these plots share a common trait: they have a mass of less than 2 M$_{\rm Jup}$ and an orbital period shorter than 1500 days. In Figure \ref{fig:statanalysis}a, the filled red circles represent all RV-detected exoplanets, while the triangles represent RV-detected exoplanets in this study. This investigation is focused on RV-detected exoplanets accompanied by debris discs and demonstrates a distribution pattern akin to previously detected RV-detected exoplanets. In Figure \ref{fig:statanalysis}a, the blue triangles align predominantly with regions that represent historically abundant areas of RV-detected exoplanets. In Figures \ref{fig:statanalysis}a,b, and c we use black dashed lines for the area occupied by hot Jupiters and warm Jupiters. Hot Jupiters are planets with a short orbit, less than 10 days, and a planet mass greater than 0.1 M$_{\rm Jup}$ \citep{yee2021complete,wright2012frequency} and have an occurrence rate ranging from  0.4$-$1.5 $\%$ \citep{beleznay2022exploring, kunimoto2020searching, petigura2018california, deleuil2018planets, fressin2013false, howard2012planet, mayor2011harps, cumming2008keck}. Warm Jupiters also have a planet mass greater than 0.1 M$_{\rm Jup}$ but with orbits from 10 to 100 days \citep{wu2018ttv, huang2016warm}. \citet{dawson2018origins} suggests that warm Jupiters are less common than hot Jupiters. In our study, no exoplanets were detected within the demarcated region outlined by dashed lines in Figure \ref{fig:statanalysis}a until the correction for sin $i$ is made in Figure \ref{fig:statanalysis}b and c.\\ 
\\
In Figure \ref{fig:statanalysis}c, the circles represent data extracted from the NASA Exoplanet Archive and include all exoplanets whose true masses have been determined either by a combination of the transit method and RV method, or by using the astrometric method and RV method. The triangles represent the $M_{DA}$ of the RV-detected exoplanets in this study. The observations are color-coded based on the stellar distances of the stars around which the exoplanets are detected. From the figure, it is apparent that the triangles represent exoplanets at much closer distances. So although the exoplanets detected around stars with debris discs are relatively few, they are among the closest examples of exoplanets with `known' masses. It should be emphasized that the relatively high concentration of hot and warm Jupiter's in this plot arises due to the high sensitivity of transit surveys to detect such signals and not on their relative abundance.\\
\\
We particularly note the three objects marked with blue triangles in Figure \ref{fig:statanalysis}b and c: HD 69830 b, HD 69830 c, and 61 Vir b. These objects have M$_{DA}$ values that place them within the designated region for hot and warm Jupiters. This suggests that finding the orbital inclination angles of more RV-detected exoplanets could reveal additional planets in the hot and warm Jupiter zones. Additionally, \citet{lin1996orbital} proposed that hot Jupiters cannot form in situ but must migrate from the cold, icy regions of the protoplanetary disc, several astronomical units from the star. Since debris discs originate from protoplanetary discs, the detection of three planets in the hot and warm Jupiter zone may indicate nearby environments with relatively recent type-II disc migration. This highlights these planets as targets for studying planet migration in protoplanetary discs \citep{alibert2005models}.\\
\\

\section{Acknowledgement}
The utilization of Python 3 and Jupyter Notebook \citep{kluyver2016jupyter} has facilitated efficient and effective progress. Special gratitude is extended to the DACE platform for its exceptional functionality in the analysis of RV data and the extraction of exoplanetary parameters, significantly amplifying the quality and depth of our results. Finally, we would like to thank the referee for their valuable feedback, which has greatly improved the clarity of the paper and whose probing comments have served to reveal important aspects of this work.

\section{Data Availability}
All the data used for this research is based on publically available data. The information regarding debris discs is accessible through The Catalog of Circums0009-0006-7546-5402tellar Discs, which can be accessed at: \url{https://www.circumstellardisks.org/}. The Transit Exoplanet Survey Satellite (TESS) data employed in this study is available through the Mikulski Archive for Space Telescopes (MAST), accessible at: \url{https://archive.stsci.edu/missions-and-data/tess}. The RV data utilized in our study is derived from various repositories:
\begin{itemize}
    \item The ESO archive, hosted by the European Southern Observatory, provides data at: \url{https://www.eso.org/public/}.
    \item The Data Analysis Center for Exoplanets (DACE) project at the University of Geneva offers RV data through their Python API, accessible at: \url{https://dace.unige.ch/pythonAPI/?tutorialId=21}.
    \item The HARPS-RVBank archive, curated by the Max-Planck-Institut für Astronomie, can be accessed at: \url{https://www2.mpia-hd.mpg.de/homes/trifonov/HARPS_RVBank.html}.
    \item Exostriker, a project hosted on GitHub by Trifon Trifonov, MPIA Heidelberg, provides RV data at: \url{https://github.com/3fon3fonov/exostriker}.
\end{itemize}
Lastly, historical exoplanet orbital parameter data is obtained from the NASA Exoplanet Archive, hosted by the California Institute of Technology. This archive can be accessed at: \url{https://exoplanetarchive.ipac.caltech.edu/}, and is operated under contract with the National Aeronautics and Space Administration (NASA).

\bibliographystyle{mnras}
\bibliography{main} 




\bsp	
\label{lastpage}
\end{document}